\documentstyle[12pt]{report}

\makeatletter

\def\thebibliography#1{\chapter*{References\@mkboth
 {Refernces}{References}}\list
 {[\arabic{enumi}]}{\settowidth\labelwidth{[#1]}\leftmargin\labelwidth
 \advance\leftmargin\labelsep
 \usecounter{enumi}}
 \def\newblock{\hskip .11em plus .33em minus -.07em}
 \sloppy
 \sfcode`\.=1000\relax}

\if@twoside
\else \oddsidemargin 00pt \evensidemargin 00pt
\fi
\def\chapter{\@startsection {chapter}{0}{\z@}{+6.0ex plus +1ex minus
 +.2ex}{2.8ex plus .2ex}{\large\bf}}
\def\section{\@startsection {section}{1}{\z@}{+3.0ex plus +1ex minus
 +.2ex}{2.3ex plus .2ex}{\normalsize\bf}}
\def\subsection{\@startsection{subsection}{2}{\z@}{+2.5ex plus +1ex
minus +.2ex}{1.5ex plus .2ex}{\normalsize\bf}}
\def\subsubsection{\@startsection{subsubsection}{3}{\z@}{+3.25ex plus
 +1ex minus +.2ex}{1.5ex plus .2ex}{\normalsize\bf}}

\def\appendix{\par
 \setcounter{chapter}{0} \setcounter{section}{0}
 \def\@chapapp{Appendix}
 \def\thechapter{\Alph{chapter}}}

 \def\wick#1{\mathop{\vtop{\ialign{##\crcr
  $\hfil\displaystyle{#1}\hfil$\crcr\noalign{\kern1\p@\nointerlineskip}
 \undercontractfill\crcr\noalign{\kern-5\p@}}}}\limits}
 \def\undercontractfill{$\m@th\rule{2pt}{0pt}\rule[2.6pt]{0.5pt}{3pt}
 \rule{-1pt}{0pt}\mathord-%
  \leaders\hbox{$\mkern-13.5mu\mathord-$}\hfill
   \mathord{}\rule{-1.5pt}{0pt}\rule[2.6pt]{0.5pt}{3pt}\rule{3pt}{0pt}$}
\expandafter\ifx\csname mathrm\endcsname\relax\def\mathrm#1{{\rm #1}}\fi

\makeatletter

\begin{document}

\newcommand{\ma}{{\em Mathematica\/} }
\newcommand{\maX}{{\em Mathematica\/}}
\newcommand{\fea}{{\em FeynArts\/} }
\newcommand{\feax}{{\em FeynArts\/}}
\newcommand{\fec}{{\em FeynCalc\/} }
\newcommand{\fecx}{{\em FeynCalc\/}}
\newcommand{\two}{{\em TwoCalc\/} }
\newcommand{\twox}{{\em TwoCalc\/}}
\hyphenation{TwoCalc}
\hyphenation{FeynArts}
\hyphenation{FeynCalc}
\hyphenation{MATHEMATICA}
\newcommand{\no}{\nonumber \\ }
\newcommand{\sun}{\circ \hspace{-0.41em} \vert \hspace{0.4em}}
\newcommand{\kett}{\bigcirc \hspace{-0.65em}\vert \hspace{0.7em}}
\newcommand{\thc}{\theta \hspace{-0.2em}/\hspace{0.1em}}
\newcommand{\as}{a\hspace{-0.5em}/\hspace{0.1em}}
\newcommand{\es}{\varepsilon \hspace{-0.5em}/\hspace{0.1em}}
\newcommand{\ks}{k\hspace{-0.52em}/\hspace{0.1em}}
\newcommand{\ps}{p\hspace{-0.42em}/\hspace{0.1em}}
\newcommand{\qs}{q\hspace{-0.5em}/\hspace{0.1em}}
\newcommand{\hb}{h\hspace{-0.5em}\vspace{-0.3em}-\hspace{0.1em}}
\newcommand{\rn}{\mbox{${\cal R}^{n}$}}
\newcommand{\rpn}{\mbox{${\cal RP}_{n-1}$}}
\newcommand{\cn}{\mbox{${\cal C}^{n}$}}
\newcommand{\cpn}{\mbox{${\cal CP}^{n-1}$}}
\newcommand{\set}[1]{\mbox{$\cal #1$}}
\newcommand{\lie}{\mbox{{\it \char'044}}}
\newcommand{\con}{\mbox{\setlength{\unitlength}{1mm}\begin{picture}(2.7,3.2)
\thicklines
\put(0.3,0.1){\line(1,0){1.7}}
\put(2.0,0.1){\line(0,1){3.0}}
\end{picture}}}

\newcommand{\scon}{\mbox{\setlength{\unitlength}{1mm}\begin{picture}(2.1,2.5)
\thicklines
\put(0.3,0.1){\line(1,0){1.2}}
\put(1.5,0.1){\line(0,1){2.1}}
\end{picture}}}

\newcommand{\la}{\langle}
\newcommand{\ra}{\rangle}
\newcommand{\dla}{\langle \langle}
\newcommand{\dra}{\rangle \rangle}
\newcommand{\noI}{\noindent}
\setcounter{page}{0}
\thispagestyle{empty}

\begin{center}
\noindent
\mbox{}\\
\vspace{1.0 cm}
{\large \bf Reduction of general two-loop self-energies
to\\
standard scalar integrals\footnote{
Accepted for publication in Nuclear Physics B} \par} \vskip 2.5em
{
{\sc G.~Weiglein\footnote{E-mail address:
weiglein@vax.rz.uni-wuerzburg.d400.de}, R.~Scharf,
M.~B\"ohm} \\[1ex]
{\it Physikalisches Institut der Universit\"at W\"urzburg,\\
Am Hubland, D-97074 W\"urzburg, Federal Republic of Germany}
\vskip 2em
February 1993
\par} \vskip 12em
\par

{\bf Abstract}
\end{center}\par \vskip 1em
{\small
\noindent
A method is presented for reducing general two-loop self-energies
to standard scalar integrals in massive gauge
theories with special emphasis on the electroweak Standard
Model (SM). We develop a technique for treating the tensor
structure of two-loop integrals appearing in self-energy
calculations. It is used together with the symmetry properties
of the integrals to obtain a result in terms of a
small number of standard scalar integrals.
The results are valid for arbitrary values of the invariant
momentum $p^2$, all particle masses, the space-time dimension
$D$ and the gauge parameters $\xi _i \; (i = \gamma , Z, W)$.
The algebraic structure of the results clearly displays the
gauge dependence of the considered quantities and allows to
perform very stringent checks. We explicitly verify
Slavnov-Taylor identities by calculating several thousand
Feynman-diagrams and adding them up algebraically.
As an application of our algorithm we calculate the light fermion
contributions to the two-loop gauge boson self-energies of the
electroweak SM. We study their gauge dependence and
discuss the occurring standard integrals.
}
\newpage

\chapter{Introduction}

The $e^{+} e^{-}$ colliders LEP100 and SLC started a new era of
precision measurements which allow to test the electroweak
Standard Model (SM) on its quantum level. In order to match the
experimental precision radiative
corrections have to be incorporated into the theoretical
predictions.
During the last years many calculations of electroweak virtual
corrections have been carried out at the one-loop level.
For the high precision experiments at LEP100, however,
first order corrections alone are inadequate.
Leading second order effects are often taken into account by
means of renormalization group methods, but rather limited
results have been obtained for irreducible virtual two-loop
corrections.

In the study of these contributions the
self-energies play a central role. These  so-called oblique
corrections are universal, i.e.~process-independent, in contrast
to the process-specific contributions due to vertices, boxes and
bremsstrahlung. A number of
authors~\cite{oblique,KenLy} recently stressed the
importance of the oblique corrections for the analysis of
precision experiments and in reformulations of electroweak
radiative corrections using effective lagrangians. In view of
the measurements at LEP100, in particular a precise calculation of
the Z-boson self-energy is of interest.

In the SM so far no complete calculation of a two-loop
self-energy has been carried out. This fact
is  due to the complicated
structure and large number of the Feynman diagrams contributing
at the two-loop level.
Results were obtained treating the limiting cases of a heavy fermion
doublet~\cite{vdBH}, a heavy top quark~\cite{barb2} and a large
Higgs mass~\cite{vdBV}. The result given
in~\cite{vdBH} was used in~\cite{ConHol} for studying the
resummation of effects due to fermion doublets with large mass
splitting.

In view of the fact that the mass expected for the top-quark is
of the order of the heavy gauge boson masses and that almost no
restrictions can be imposed on the Higgs mass
a calculation
allowing for general values of the top and Higgs mass, the gauge
boson masses and the invariant momentum $p^2$ might be of
interest.

In this paper we present a systematic way for treating all two-loop
self-energies. It is applicable for gauge bosons, scalars,
gauge boson-scalar mixing and fermions.
In our discussion we focus on
the electroweak SM, but the method is valid for any
renormalizable model.
The strategy we have adopted is to reduce the
amplitudes algebraically as far as possible.
We develop a technique for the tensor decomposition of two-loop
self-energy integrals.
This generalizes the results for one-loop integrals
worked out by Passarino and Veltman~\cite{pass}.
Using this technique and the symmetry properties
of the integrals we obtain a result in which the Feynman
amplitudes are given in terms of standard scalar two-loop
integrals.
This is in analogy to one-loop calculations where the Feynman
amplitudes are expressed in terms of the standard
integrals $A_0,B_0,C_0$ and $D_0$ defined in~\cite{thove}.
We will show that the reduction to the specified class of integrals is
possible for every two-loop self-energy.

The calculations are performed for arbitrary values of
all particle masses, the invariant momentum $p^2$ and the
space-time dimension $D$. We work in a general $R_{\xi}$-gauge
specified by one gauge parameter $\xi _i \; (i = \gamma , Z, W)$
for each vector boson. This allows to study the gauge dependence
of the calculated quantities.
The gauge dependence of basic electroweak
corrections has recently found considerable interest and was
studied at one-loop order by a number of
authors~\cite{KenLy,gauge1,pinch,Sirlin}.
At the two-loop level this issue is of
even greater importance since as long as the calculation of a
complete process, which necessarily is gauge invariant, is out
of reach it is crucial to know the gauge
dependence of the results.
We will show that our method of
representing the results in terms of two-loop standard integrals
very clearly displays the gauge dependence of the considered
quantities. It can be read off directly from the algebraic
result. There is no need for using an explicit analytical or
numerical expression of the standard integrals.

We show that our algorithm for treating the two-loop
self-energies is well suited for calculations involving a large
number of Feynman diagrams.
The results consist of a relatively small number of standard integrals
which can conveniently be studied for further evaluation. In
contrast to that a direct evaluation of the tensor integrals
would in general involve a large number of integrals,
each one to be treated separately.
The results we obtain are transparent and have the
benefit that very stringent checks can easily be performed on them.
In analogy to the investigation of the gauge dependence
we can check Slavnov-Taylor identities directly at the
algebraic level. This test is exact, i.e.~free of any
numerical uncertainty.
We implemented our algorithm in the computer-algebra program
\two \cite{two}. In order to check its reliability we
explicitly verify Slavnov-Taylor identities for the
self-energies of the $ \gamma Z$-system involving
several thousand Feynman diagrams. Every graph is
calculated separately and the results are summed up algebraically.

As an application we treat the light fermion contributions to
the two-loop gauge boson self-energies in the SM. Considering
the results obtained in one-loop order the light fermions
are expected
to yield a significant contribution to the complete result. We
give the results for the $\gamma$, $Z$ and $W$ self-energies in
terms of two-loop standard integrals for general values of
$p^2$, the gauge boson masses and the Higgs mass. We
study the gauge dependence of these amplitudes and classify them in
several subsets according to their behavior under gauge
transformations.
All standard integrals appearing in the results
of the light fermion contributions can be
solved analytically leading to an expression in terms of
polylogarithmic functions. We worked this out explicitly. The
results will be presented in a related paper~\cite{raibas}.

The paper is organized as follows: In sect.~2 we classify the
two-loop self-energies according to their topologies and define
the relevant quantities. Sect.~3 is concerned with the symmetry
properties of the two-loop integrals. In sect.~4 we develop a
technique for the tensor decomposition of two-loop integrals and
show that it is applicable for every integral which arises in
the calculation of two-loop self-energies. In sect.~5 we consider
relations used for minimizing the number of occurring standard
integrals. Sects.~6 and 7 deal with the computer-algebraic
realization of the algorithm and its use for verifying
Slavnov-Taylor identities in the electroweak SM. In sect.~8 we
give results for the two-loop photon self-energy in QED and
for the light fermion contributions to the two-loop gauge boson
self-energies in the electroweak SM.
The properties of these results are
discussed. In the appendix we give a result for the $W$
self-energy and list several reduction formulae
needed for the calculations performed in this paper.
\chapter{Classification of two-loop self-energies}
\label{classif}

The two-loop self-energies can be classified according to the
topologies of the corresponding Feynman diagrams. All self-energy
topologies which can occur in
renormalizable gauge theories are shown in fig.~\ref{fig:top}.
We have listed the one-particle irreducible topologies, where
as usual diagrams containing tadpole lines are included.
Furthermore
nine reducible topologies exist which correspond to products of
one-loop self-energies. In the following we will call the first eight
topologies in fig.~\ref{fig:top} ``generic'' two-loop topologies.

\begin{figure}
\begin{picture}(350,532)
\end{picture}
\caption{The 20 topologies possible for two-loop self-energies
in renormalizable gauge theories.}
\label{fig:top}
\end{figure}

We focus on  the unrenormalized self-energies, a proper
renormalization can be done after the algebraic calculation has
been carried out. In order to make the integrals mathe\-mati\-cally
meaningful we use dimensional regularization and work in an
arbitrary space-time dimension $D$.

Inserting fields into the topologies and applying the Feynman
rules for propagators and vertices leads to the Feynman
amplitudes.
It is convenient to deal with scalar quantities,
i.e.~to begin with a tensor decomposition.
For the gauge boson
self-energies it reads
\begin{equation}
\Sigma _{\mu \nu}^{\alpha , \beta }(p) = \Bigl(-g_{\mu \nu }
+ \frac {p_{\mu}
p_{\nu}}{p^2} \Bigr) \Sigma _{T}^{\alpha , \beta }(p^2) - \frac {p_{\mu}
p_{\nu}}{p^2} \Sigma _{L}^{\alpha , \beta }(p^2) \;,  \label{eq:stens1}
\end{equation}
from which the transverse part $\Sigma _{T}^{\alpha , \beta }(p^2)$
and the
longitudinal part $\Sigma _{L}^{\alpha , \beta }(p^2)$ can easily be
extracted:
\begin{equation}
 \Sigma _{T}^{\alpha , \beta }(p^2) = \frac{1}{D-1}
 \Bigl(-g^{\mu \nu } + \frac {p^{\mu} p^{\nu}}{p^2} \Bigr)
 \Sigma _{\mu \nu}^{\alpha , \beta }(p) \; ; \; \; \Sigma
 _{L}^{\alpha , \beta }(p^2) = - \frac {p^{\mu} p^{\nu}}{p^2}
 \Sigma _{\mu \nu}^{\alpha , \beta }(p) \; , \label{eq:stens1a}
\end{equation}
$p$ is the external momentum, $D$ the space-time dimension,
$\alpha, \beta  = \gamma , Z$ for the $(\gamma
Z)$-system and $\alpha = W,$ $\, \beta  = W$ for the $W$-boson.

We write for the mixing of gauge bosons and unphysical
Higgs-fields
\begin{equation}
\Sigma _{\mu }^{\alpha ,i}(p) = p_{\mu } \Sigma ^{\alpha
,i}(p^2) \; . \label{eq:stens2}
\end{equation}
Consequently we have
\begin{equation}
\Sigma ^{\alpha ,i}(p^2) = \frac {p^{\mu}}{p^2} \Sigma _{\mu
}^{\alpha ,i}(p) \; , \label{eq:stens2a}
\end{equation}
where $i = \chi , \varphi$ for the $(\gamma Z)$-system and the
$W$-boson, respectively.

The fermion self-energies can be decomposed into a vector, an
axial vector, a scalar and a pseudoscalar part according to
\begin{equation}
\Sigma (p) = \not \! p \, \Sigma_V (p^2) + \not \! p \gamma _5
\, \Sigma_A (p^2) +  m \Sigma_S (p^2) + m \gamma _5 \Sigma_P
(p^2) \; , \label{eq:stens3}
\end{equation}
where $m$ is the mass of the fermion and mixing effects have been
suppressed for simplicity of notation. We obtain
\begin{equation}
 \Sigma_V (p^2) = \frac{1}{4 p^2} \mbox{Tr} \Bigl(\not \! p \,
 \Sigma (p)
 \Bigr) \; ; \; \; \Sigma_A (p^2) = \frac{1}{4 p^2} \mbox{Tr}
 \Bigl(\gamma _5 \! \! \not \! p \, \Sigma (p) \Bigr)
\label{eq:stens3a1}
\end{equation}
and
\begin{equation}
 \Sigma_S (p^2) = \frac{1}{4 m} \mbox{Tr} \Bigl(\Sigma (p) \Bigr)
 \; ; \; \;
 \Sigma_P (p^2) = \frac{1}{4 m} \mbox{Tr} \Bigl( \gamma _5
 \Sigma (p) \Bigr)
 \; . \label{eq:stens3a2}
\end{equation}

Adding the inverse of the zeroth order propagator to the
self-energies of all orders (including the tadpole amplitudes)
leads to the truncated one-particle irreducible two-point functions.
For the Higgs field being a Lorentz scalar this yields
\begin{equation}
 \Gamma^H(p) = i (p^2 - m_H^2) + i \Sigma^H(p^2)
\end{equation}
and for the Lorentz tensors we use the decompositions specified
above.
The corresponding propagators are obtained as the inverse of these
two-point functions.
For the neutral gauge bosons we have to consider matrices.
The transverse part of the inverse propagator matrix reads
\begin{equation}
 D_{T}^{-1} = i
 \left( \begin{array}{ll}
	 p^2 + \Sigma ^{\gamma \gamma }_{T}(p^2) &
	 \qquad \Sigma ^{\gamma Z}_{T}(p^2) \\
				       &       \\
         \Sigma ^{\gamma Z}_{T}(p^2) &
	 \qquad p^2 - M_{Z}^2 + \Sigma ^{Z Z}_{T}(p^2)
        \end{array}
 \right) \; , \label{eq:matr}
\end{equation}
from which the propagators
\begin{equation}
D_{T} = \left(
\begin{array}{ll}
 \Delta ^{\gamma \gamma }_{T} & \qquad \Delta ^{\gamma Z}_{T} \\
			      & \qquad                          \\
 \Delta ^{\gamma Z}_{T}      & \qquad \Delta ^{Z Z}_{T}
\end{array}
\right)
\end{equation}
follow by matrix inversion.


We recall that in the $R_{\xi}$-gauges the lowest order
gauge boson propagators can be written as
\begin{equation}
 D^{i}_{\mu \nu}(k) = \frac{-i g_{\mu \nu }}{ \Bigl[ k^2 - m_i^2
 \Bigr]}
 + \frac{i (1 - 1/ \xi_i )
 k_{\mu} k_{\nu}}{ \Bigl[ k^2 - m_i^2 / \xi_i \Bigr]  \Bigl[ k^2 -
 m_i^2 \Bigr] } \; , \label{eq:gabopr}
\end{equation}
where $i = \gamma , Z, W$. The parameters $\xi_i$
associated with these fields specify the gauge. They
can be chosen independently.
The 't Hooft-Feynman gauge is realized by setting $\xi_i = 1$,
$\xi_i\rightarrow \infty$ defines the Landau gauge while the
limit $\xi_i \rightarrow 0$ corresponds to the unitary gauge.
The unphysical scalar and ghost propagators are of the form
$i \Bigl[k^2 - m_i^2 / \xi_i \Bigr]^{-1}$. The fermion propagators
can be written as $i (\not \! k + m )
\Bigl[k^2 - m^2 \Bigr]^{-1}$.

In the following we will use the shorthand notation
\begin{equation}
 \langle \langle \ldots \rangle \rangle = \int \frac{d^D q_1}{i
 \pi ^2 (2 \pi \mu)^{D - 4}} \int \frac{d^D q_2}{i \pi ^2 (2 \pi
 \mu)^{D - 4}} ( \ldots )  \; , \label{eq:bez2}
\end{equation}
where $q_1$ and $q_2$ are the integration momenta of the loop
integrals and $\mu $ is an arbitrary reference mass.

The Feynman amplitudes we are concerned with can therefore be
written as
\begin{equation}
\langle \langle \frac {\cdots}{
\Bigl[k_{1}^2 - m_1^2 \Bigr] \Bigl[k_{2}^2 - m_2^2 \Bigr]
\cdots \Bigl[k_{\ell}^2 - m_\ell^2 \Bigr]}
\rangle \rangle \; , \label{eq:den}
\end{equation}
where $k_{j}$ is the momentum of the $j$-th propagator
and $m_j$ its mass, $j = 1, \ldots , \ell$. The $m^2 / \xi_i$-terms
occurring in the propagators given above are simply treated as mass
parameters. In~(\ref{eq:den}) it is understood that the masses
carry a small negative imaginary part.

The numerator of the Feynman amplitude in general has a
complicated structure being a function of the two integration
momenta, the external momentum $p$, the particle masses and the
gauge parameters $\xi_i$.
It will be a central issue of this paper to show how
all Feynman amplitudes
can be reduced to a form where the numerator consists
only of quantities which are independent of the integration
momenta and can therefore be pulled out of the integral.
The denominator is still of the form~(\ref{eq:den}).
This means that the
amplitudes can be expressed in terms of a class of scalar
two-loop integrals
\begin{equation}
T_{i_1 i_2 \ldots i_\ell} (
p^2;m_1^2,m_2^2,\ldots,m_{\ell}^2 )
= \langle \langle
\frac{1}{ \Bigl[k_{i_1}^2 - m_1^2 \Bigr] \Bigl[k_{i_2}^2 -
m_2^2 \Bigr] \cdots \Bigl[k_{i_\ell}^2 - m_\ell^2 \Bigr] }
\rangle \rangle \; , \label{eq:Tint}
\end{equation}
which we call $T$-integrals.
This is reminiscent of the well known result that
every one-loop amplitude can be reduced to the basic scalar
integrals $A_0, B_0, C_0$ and $D_0$~\cite{pass,thove}.
It should be noted, however, that it is by far not obvious that
such a reduction is possible in the general
two-loop case and we only
claim it for self-energies.

We used the double index notation $T_{i_1 i_2 \ldots i_\ell}$ to
indicate that the
subindices of the $T$-integrals refer to the corresponding
momenta $k_{i_1}, k_{i_2}, \ldots k_{i_\ell}$.
The masses are only explicitly written as
arguments if confusion is possible.
If a propagator has mass zero, we indicate this
with a prime at the corresponding subindex and drop the zero in the
list of arguments, e.g.
\begin{equation}
 T_{1'234} ( p^2;m_a^2,m_b^2,m_c^2 ) = \langle
 \langle \frac{1}{ k_1^2 \Bigl[k_2^2 - m_a^2 \Bigr] \Bigl[k_3^2
 - m_b^2 \Bigr] \Bigl[k_4^2 - m_c^2 \Bigr] } \rangle \rangle \;
 .
\end{equation}
The dependence on $p^2$ will
be suppressed in the following.

The topologies listed in fig.~\ref{fig:top} can also be used
to represent
the $T$-integrals.
A line
in the topology carrying zero momentum contributes only a factor
$ (-1/m_i^2) $ which is irrelevant for the
scalar integral. The topologies 6 and 7 in fig.~\ref{fig:top}
therefore represent the same type of $T$-integral.
The first two topologies in fig.~\ref{fig:top} are the most
general ones since all scalar integrals corresponding to the
other topologies can be obtained from these by omitting factors
$ \Bigl[k_{j}^2 - m_j^2 \Bigr] $ in the
denominator or equivalently shrinking
the corresponding lines in the topology to a point. The
integral corresponding to the first topology, being the only
one where five propagators with
different momenta occur, is
sometimes called the ``master integral''
\begin{equation}
T_{1 2 3 4 5} = \langle \langle \frac{1}{\Bigl[k_1^2 - m_1^2\Bigr]
\Bigl[k_2^2
- m_2^2\Bigr]
\Bigl[k_3^2 - m_3^2\Bigr]
\Bigl[k_4^2 - m_4^2\Bigr]
\Bigl[k_5^2 - m_5^2\Bigr]}
\rangle \rangle \; ,
\end{equation}
where the momenta are labeled as indicated in fig.~\ref{fig:top1}.
\begin{figure}
\begin{picture}(450,70)
\end{picture}
\caption{The topology of the ``master integral''}
\label{fig:top1}
\end{figure}
In the following we do not manifestly impose momentum conservation,
but keep the overcomplete set of momenta $k_1 ,
\ldots , k_5$. This will be convenient especially
for treating the symmetry properties of the integrals and
performing the tensor reductions. It is easy to reexpress
the momenta $k_1 , \ldots , k_5$ by the external momentum
$p$ and the integration momenta
$q_1$ and $q_2$:
\begin{equation}
k_1 = q_1, \; k_2 = q_1 + p, \; k_3 = q_2 - q_1, \; k_4 = q_2,
\; k_5 = q_2 + p. \label{eq:momenta}
\end{equation}
With this convention, the integral corresponding to the second
topology in fig.~\ref{fig:top}, for example, can be written
as $T_{11234}$. We will see below that equivalent representations
can be obtained by certain permutations of the indices.
\chapter{Symmetries of the two-loop integrals}

As stated above we start with a tensor decomposition
(eqs.~(\ref{eq:stens1}),~(\ref{eq:stens2})
and~(\ref{eq:stens3}), respectively) in order to obtain
scalar quantities.
The contraction of Lorentz indices, reduction of the Dirac
algebra and evaluation of Dirac traces can be worked out like in
the one-loop case. This results in scalar products of
momenta $(k_i \cdot k_j),$ $\, (k_i \cdot p),$ $\, p^2$ in the
numerator of the Feynman amplitude. The
denominator is unchanged. We now implicitly use momentum
conservation and express all scalar products as sums of momentum
squares, e.g.
\begin{equation}
(k_1 \cdot p) = \frac {1}{2} (k_2^2 - k_1^2 - p^2) \; .
\end{equation}
Subsequently all $k_i^2$ appearing both in the numerator and
the denominator are canceled via
\begin{equation}
k_i^2 = (k_i^2 - m_i^2) + m_i^2 \; .
\end{equation}
The application of this procedure directly leads to
$T$-integrals, if all $k_i^2$ appearing in the numerator can be
canceled. If not, we obtain another type of integral which contains
squares of momenta in the numerator not occurring in the
denominator:
\begin{equation}
 Y_{j k l \ldots} ^{i \ldots}  =  \langle \langle
 k_{i}^2 \cdots \;
 \Delta_{j k l \ldots}
 \rangle \rangle
  \; , \; \; \; i,j,k,l = 1, \ldots , 5 \; , \; \; i \neq
 j,k,l \; .
\end{equation}
Here we used the shorthand notation
\begin{equation}
 \Delta_{j k l \ldots} = \frac{1}{\Bigl[k_j^2 - m_j^2 \Bigr]
 \Bigl[k_k^2 - m_k^2 \Bigr] \Bigl[k_l^2 - m_l^2 \Bigr] \cdots} \; .
\end{equation}

We now have obtained a representation of the amplitudes in terms
of scalar $T$- and $Y$-integrals. These integrals, however, are
not independent of each other. The number of occurring integrals
can considerably be reduced by taking into account their symmetries with
respect to permutation of the $k_i^2$ or, equivalently, of the
corresponding indices.

All $T$- and $Y$-integrals are invariant under the permutations
\begin{equation}
 (12)(45), \; \; (14)(25), \; \; (15)(24). \label{eq:perm1}
\end{equation}
Here, as always, $k_1 , \ldots , k_5$
are labeled according
to~(\ref{eq:momenta}).
Application of the first permutation yields for example
\begin{eqnarray}
T_{1 2 3 4}(p^2; m_{a}^2, m_{b}^2, m_{c}^2, m_{d}^2) & = &
T_{1 2 3 5}(p^2;
m_{b}^2, m_{a}^2, m_{c}^2, m_{d}^2) \no
& = & \langle \langle \frac{1}{\Bigl[k_1^2 - m_b^2 \Bigr]
\Bigl[k_2^2 - m_a^2 \Bigr]
\Bigl[k_3^2 - m_c^2 \Bigr]
\Bigl[k_5^2 - m_d^2 \Bigr]} \rangle \rangle \; .
\end{eqnarray}

The validity of the symmetries listed
in~(\ref{eq:perm1}) can most easily be seen for the
``master integral'' (fig.~\ref{fig:top1}) or can be checked
using~(\ref{eq:momenta}) and the invariance properties of the
integrals with respect to changes of the integration momenta.

Additional symmetry relations hold if an index does
not occur in the integral, e.g.
\begin{equation}
 (2 3) \; \mbox{if 1 is absent}.  \label{eq:perm2}
\end{equation}
The symmetry relations~(\ref{eq:perm1}) and~(\ref{eq:perm2}) are
used to map every integral onto a standard representative,
i.e.~all integrals which are related by symmetry transformations
are brought to the same form.
\chapter{Tensor reduction for two-loop integrals}

As claimed above, all self-energy amplitudes can be expressed by
an independent set of $T$-integrals. This means that all
$Y$-integrals can further be reduced. To achieve this, we
rewrite the $Y$-integrals in terms of tensor integrals and
perform a tensor decomposition.

As a simple example we consider the integral
\begin{equation}
 Y_{2 3 4 5} ^1 = \langle \langle k_{1}^2 \Delta _{2 3 4 5}
 \rangle \rangle \label{eq:1Y2345} \; .
\end{equation}
Insertion of
\begin{equation}
 k_{1}^2 = (k_{2}^2 - m_{2}^2) + (m_2^2 + p^2) - 2 (p \cdot k_{2})
\end{equation}
yields
\begin{equation}
 Y_{2 3 4 5} ^1 = T_{3 4 5} + ( m_2 ^2 + p^2 ) T_{2 3 4 5} - 2
 p_{\mu} \langle \langle k_{2} ^{\mu} \Delta _{2 3 4 5} \rangle
 \rangle \; . \label{eq:63}
\end{equation}
Now one has to perform a tensor decomposition for the integral
\begin{equation}
S_{2 3 4 5} ^{2  , \; \mu} = \langle \langle k_{2} ^{\mu}
\Delta _{2 3 4 5} \rangle \rangle \; . \label{eq:te2345}
\end{equation}
It is obvious that the Passarino-Veltman procedure~\cite{pass} used for
one-loop integrals, i.e.~the ansatz
\begin{equation}
S_{2 3 4 5} ^{2  , \; \mu} = p^{\mu} S(p^2) \;,
\label{eq:pas}
\end{equation}
does not lead to simpler integrals in this case. To determine
the scalar quantity $S(p^2)$ one has to contract with $p_{\mu}$,
but it is not possible to cancel the resulting scalar product
$(p \cdot k_2)$, since it is not expressible as a sum of squared momenta
occurring in the denominator of the integral. This is due to the
fact that the momenta $p$ and $k_2$ belong to a four-vertex in
the topology
$\Delta _{2 3 4 5}$ (see
fig.~\ref{fig:top2}). This is a typical feature of two-loop
topologies, whereas in one-loop order only three-vertices occur
in the loop integral and the scalar products can always be
canceled.

In our approach we not only work with the tensor structure of the
integral with respect to the external momentum $p$ but also perform
decompositions with respect to a subloop.
In general one can write for
a subloop $s^{\mu}$ which has the structure of a first-rank tensor
\begin{equation}
 s^{\mu} = p_1^{\mu} s_1 + p_2^{\mu} s_2 + \ldots
 \label{eq:tensallg}
\end{equation}
where $s_1, s_2, \ldots$ are Lorentz-scalars and
$p_1, p_2, \ldots$ are the independent external momenta
of the subloop.

At first sight it is not obvious that the
scalar products resulting from this decomposition can be
canceled. Moreover, as is well known from one-loop calculations,
contracting~(\ref{eq:tensallg}) with $p_{1 \mu}, p_{2 \mu},
\ldots$ and solving for $s_1, s_2, \ldots$ leads to factors which
are determinants of the momenta $p_i$. These, however, depend on
the integration momentum of the second loop and cannot be pulled
out of the tensor integral like in the one-loop case.

In order to show how
this technique can successfully be applied to all two-loop
self-energies we first examine, which
denominators of tensor integrals can occur after the steps
performed in the previous section. They can be read off
from the topologies listed in fig.~\ref{fig:top}.
Topology 1 clearly is not possible, since at
least one of its five different propagators can always be
canceled. The
topologies 9
through 20 are just products of one-loop contributions. For
these the Passarino-Veltman technique can be applied in a
straightforward way. The tensor decomposition for the topologies
6 through 8, having no dependence on the external momentum $p$,
is also trivial.

We therefore have to consider the topologies 2 -- 5. As shown in
fig.~\ref{fig:top2} they can be written as $\Delta _{2 3 4 5 5}$,
$\Delta _{2 3 4}$ and $\Delta _{2 3 4 5}$, respectively. Note that the
scalar integrals corresponding to topologies 4 and 5 are equivalent
due to the symmetry relations~(\ref{eq:perm1}).
The topologies we have to deal with all contain at least
one subloop which is a self-energy insertion, in our terminology
this is $\Delta _{2 3}$. We do the decomposition for this
subloop. In the simplest case of a first-rank tensor we have
\begin{equation}
 \langle k_{2} ^{\mu} \Delta _{2 3} \rangle = k_{5} ^{\mu} \,
 s(k_5^2)
 \label{eq:decomp}
\end{equation}
which is just the example considered in~(\ref{eq:te2345}).
$k_{5}$ is
the external momentum of the subloop $\Delta _{2 3}$. The
Lorentz-scalar $s(k_5^2)$ is obtained from
\begin{equation}
s(k_5^2) = \frac{1}{k_{5}^2} \langle (k_5 \cdot k_2) \Delta _{2 3}
\rangle = \langle (k_5 \cdot k_2) \Delta _{2 3 5'} \rangle
\; . \label{eq:PASS1}
\end{equation}
Due to the self-energy structure of $\Delta _{2 3}$
the determinant $1/k_{5}^2$ has the form of a massless
propagator. Insertion of~(\ref{eq:PASS1}) into~(\ref{eq:decomp})
and~(\ref{eq:te2345}) therefore leads to integrals which still
belong to the class of $T$- and $Y$-integrals. We get
\begin{equation}
p_{\mu} S_{2 3 4 5} ^{2  , \; \mu}
= \langle \langle (p \cdot k_5) (k_5 \cdot k_2) \Delta _{2
3 4 5 5'} \rangle \rangle \; . \label{eq:reins}
\end{equation}
Since the momenta $k_5$ and $k_2$ belong to a three vertex in
the topology $\Delta _{2 3 4 5}$ their scalar product is
expressible as a sum of squared momenta which can be canceled:
\begin{equation}
 (k_5 \cdot k_2) = \frac{1}{2} (k_2^2 - k_3^2 + k_5^2) \; .
 \label{eq:prod1}
\end{equation}
The same holds for $(p \cdot k_5)$
\begin{equation}
 (p \cdot k_5) = \frac{1}{2} (k_5^2 - k_4^2 + p^2) \; ,
 \label{eq:prod2}
\end{equation}
if one notes that one of the $k_5^2$-terms can be canceled
against the massless propagator $\Delta _{5'}$.

\begin{figure}
\begin{picture}(450,110)
\put(81,10){$\Delta_{23455}$}
\put(219,10){$\Delta_{234}$}
\put(357,10){$\Delta_{2345}$}
\end{picture}
\caption{The topologies representing the denominators of tensor
integrals for which the Passarino-Veltman technique is not
directly applicable}
\label{fig:top2}
\end{figure}

We can therefore express $Y_{2 3 4 5} ^1$ solely in terms of
$T$-integrals. The result reads
\begin{eqnarray}
Y_{2 3 4 5}^1 &=& A_0(m_3^2) B_0(p^2;m_4^2,m_5^2)
 - \frac{1}{2} \Biggl\{ \Bigl[ A_0(m_2^2)-A_0(m_3^2) \Bigr] \Bigl[
\frac{1}{m_5^2} A_0(m_5^2) \no
&& - B_0(p^2;m_4^2,m_5^2) \Bigr]
 + T_{ 2 3 4}
 - T_{ 2 3 5}
- (m_2^2 + m_3^2 + m_4^2 -
m_5^2 + p^2) T_{ 2 3 4 5}  \no
&& - (m_2^2 - m_3^2) T_{ 2 3 5
5'} + (m_4^2-p^2) \Bigl[ T_{ 2 4 5 5'} - T_{3 4 5 5'} \Bigr]
\no
&& - (m_2^2 - m_3^2) (m_4^2-p^2)
T_{ 2 3 4 5 5'} \Biggr\} \; .
\end{eqnarray}
The one-loop scalar integrals $A_0$ and $B_0$ appear since
some $T$-integrals
are expressible as
products of one-loop integrals, e.g.
\begin{equation}
 T_{3 4 5} = \langle \langle \Delta _{3 4 5} \rangle \rangle =
 \langle \Delta _{3} \rangle \langle \Delta _{4 5} \rangle =
 A_0(m_3^2) \, B_0(p^2;m_4^2,m_5^2) \; ,
\end{equation}
where
\begin{equation}
 A_0(m^2) = \langle \frac{1}{q^2 - m^2} \rangle \; ,
\end{equation}
\begin{equation}
 B_0(p^2;m_1^2,m_2^2) = \langle \frac{1}{\Bigl[q^2 - m_1^2
 \Bigr] \Bigl[(q + p)^2 - m_2^2 \Bigr]} \rangle \; ,
\end{equation}
and we use
\begin{equation}
 \langle \ldots \rangle = \int \frac{d^D q}{i \pi ^2 (2 \pi
 \mu)^{D - 4}} ( \ldots )
\end{equation}
in analogy to~(\ref{eq:bez2}).
Note that our definitions slightly differ from those used
in~\cite{thove}.

The integrals containing a massive and a massless propagator
with the same momentum can further be simplified by partial
fractioning.

The topology $\Delta _{2 3 4 5 5}$ is treated in analogy to
$\Delta _{2 3 4 5}$. The corresponding integrals are easily
obtained by taking the derivative with respect to $m_5^2$, e.g.
\begin{equation}
 Y_{2 3 4 5 5} ^{1} = \frac{\partial }{\partial (m_5 ^2)} Y_{2 3 4
 5} ^{1} \; .
\end{equation}

The decomposition for the tensors of higher
rank can be worked out in a similar manner as described above.
In order to treat $Y_{2 3 4 5} ^{1 1}$ we have to
decompose the integral
\begin{equation}
 \langle \langle (p \cdot k_{2}) (p \cdot k_{2})
 \Delta _{2 3 4 5} \rangle
 \rangle = p_{\mu} p_{\nu} \langle \Delta _{4 5} \langle k_{2}
 ^{\mu} k_{2} ^{\nu} \Delta _{2 3} \rangle \rangle \; .
 \label{eq:s222345}
\end{equation}
We write
\begin{equation}
\langle k_{2} ^{\mu} k_{2} ^{\nu} \Delta _{2 3} \rangle =
s_{0 0}(k_5^2) g^{\mu \nu} + s_{1 1}(k_5^2) \frac{k_{5} ^{\mu}
k_{5} ^{\nu}}{k_5^2} \label{eq:compmn}
\end{equation}
with the Lorentz-scalars  $s_{0 0}(k_5^2)$ and $s_{1
1}(k_5^2)$. Contraction with $g_{\mu \nu}$ and $k_{5 \mu} k_{5
\nu}$, respectively, leads to two equations for the scalars
$s_{0 0}$ and $s_{1 1}$:
\begin{eqnarray}
s_{0 0} D + s_{1 1} &=& \langle k_2 ^2 \Delta _{2 3} \rangle \no
s_{0 0} + s_{1 1} &=& \frac{1}{k_5^2} \langle (k_5 \cdot k_2) (k_5
\cdot k_2) \Delta _{2 3} \rangle \; . \label{eq:tensmn}
\end{eqnarray}
Again, the factors $1/k_5^2$ appearing in~(\ref{eq:compmn})
and~(\ref{eq:tensmn}) can be written as propagators with mass
zero. Inspection of~(\ref{eq:tensmn}) and~(\ref{eq:s222345})
shows that the decomposition for the integral~(\ref{eq:s222345})
leads to $T$-integrals and one $Y$-integral $Y_{2 3 4}^5$ which
corresponds to a tensor of lower rank. The tensor reduction of
this integral will be worked out below, where we discuss the
integrals with the topology $\Delta _{2 3 4}$. The complete
reduction formula for $Y_{2 3 4 5}^{1 1}$ is rather lengthy.
We give it in the appendix.

The generalization to tensors of higher than second rank should
now be obvious. The decomposition of the tensor integrals always
leads directly to $T$-integrals or to tensor integrals of lower
rank.

So far we showed that the tensor decomposition works for all
tensor integrals with the topologies $\Delta _{2 3 4 5}$ and
$\Delta _{2 3 4 5 5}$. The third topology we have to consider is
$\Delta _{2 3 4}$. This case is slightly more complicated due to
the presence of the two four-vertices in the topology. The
$Y$-integrals which can occur are $Y_{2 3 4}^1$, $Y_{2 3
4}^5$, $Y_{2 3 4}^{1 1}$, $Y_{2 3 4}^{1 5}$, $Y_{2 3 4}^{5 5},
\ldots$. The integrals
$Y_{2 3 4}^1$ and $Y_{2 3 4}^5$ are not independent
of each other but are
related through a symmetry relation which amounts to a
permutation of the masses $m_2$ and $m_4$. The same holds, of
course, for $Y_{2 3 4}^{1 1}$ and $Y_{2 3 4}^{5 5}$.

We begin with the integral $Y_{2 3 4}^1$ and apply exactly the
same steps as for $Y_{2 3 4 5}^1$. This gives
\begin{eqnarray}
Y_{2 3 4} ^{1} & = & T_{3 4} + (m_2 ^2 + p^2) T_{2 3 4} - 2
p_{\mu} \langle \Delta _4 \langle k_2 ^{\mu} \Delta _{2 3}
\rangle \rangle \no
& = & T_{3 4} + (m_2 ^2 + p^2) T_{2 3 4} - 2 \langle \langle
(p \cdot k_5) (k_5 \cdot k_2) \Delta _{2 3 4 5'} \rangle \rangle \; ,
\end{eqnarray}
where~(\ref{eq:decomp}) and~(\ref{eq:PASS1}) were used. The last
integral looks very similar to~(\ref{eq:reins}) but here
it is not possible to cancel the $k_5^4$ term resulting
from the insertion of~(\ref{eq:prod1}) and~(\ref{eq:prod2}).
Instead we obtain
\begin{equation}
Y_{2 3 4} ^{1}(m_2^2,m_3^2,m_4^2) = - \frac{1}{2} Y_{2 3 4}
^{5}(m_2^2,m_3^2,m_4^2) + f(T) \; , \label{eq:1y234a}
\end{equation}
where $f(T)$ summarizes terms in which only $T$-integrals occur.
In order to find an expression for $Y_{2 3 4} ^{1}$
in terms of $T$-integrals only, we use the last symmetry relation
listed in~(\ref{eq:perm1}) and obtain after appropriate
relabeling of the masses
\begin{equation}
Y_{2 3 4} ^{5}(m_2^2,m_3^2,m_4^2) = - \frac{1}{2} Y_{2 3 4}
^{1}(m_2^2,m_3^2,m_4^2) + g(T) \; , \label{eq:1y234b}
\end{equation}
where $g(T)$ again depends on $T$-integrals only.
Insertion of this equation into~(\ref{eq:1y234a}) gives the
result
\begin{eqnarray}
Y_{2 3 4} ^{1} & = & \frac{1}{3} \Biggl\{ h(A_0, B_0)
+ (m_2^2 + m_3^2 + m_4^2 + p^2) T_{ 2 3 4}
+ 2 (m_2^2 - m_3^2) T_{ 2 3 5'} + (m_3^2 - m_4^2) T_{1'34} \no
& & + 2 (m_2^2 - m_3^2) (m_4^2 - p^2) T_{ 2 3 4 5'} + (m_2^2 - p^2)
(m_3^2 - m_4^2) T_{1' 2 3 4} \Biggr \}  \; ,
 \label{eq:1y234-1}
\end{eqnarray}
where $h(A_0, B_0)$ represents a function containing only one-loop
integrals.

Another complication can be seen by
inspecting~(\ref{eq:1y234-1}). Due to the absence of the
momentum $k_5$ the integral $Y_{2 3 4} ^{1}$ is symmetric with
respect to the permutation of the momenta $k_3$ and $k_4$ or,
equivalently, of the masses $m_3$ and $m_4$. However, the tensor
decomposition artificially introduces the momentum $k_5$ into
the terms on the right hand side
of~(\ref{eq:1y234-1}) and the
symmetry seems to be lost.
It can be made manifest again
by using the relation~(\ref{eq:Trel1}) stated
below. The result for $Y_{2 3 4} ^{1}$ finally reads
\begin{eqnarray}
 Y_{2 3 4}^1 &=& \frac{1}{3} \Biggl\{
 A_0(m_2^2) A_0(m_3^2)+A_0(m_2^2) A_0(m_4^2)+A_0(m_3^2) A_0(m_4^2) +
 (m_2^2 + m_3^2 + m_4^2 + p^2) T_{ 2 3 4 } \no
 &&- (m_4^2 - p^2) \Bigl[A_0(m_2^2)-A_0(m_3^2) \Bigr] B_0(p^2;m_4^2,0)
 - (m_3^2 - p^2) \Bigl[A_0(m_2^2)-A_0(m_4^2) \Bigr] \no
 && \times B_0(p^2;m_3^2,0)
 + (m_2^2 - m_3^2) T_{ 2 3 5'}
 +(m_2^2 - m_4^2) T_{2 3 5'}(m_2^2,m_4^2) \no
 &&+ (m_2^2 - m_3^2) (m_4^2 - p^2) T_{ 2 3 4 5'}
 +(m_3^2 - p^2) (m_2^2 - m_4^2)
T_{2 3 4 5'}(m_2^2,m_4^2,m_3^2) \Biggr\} \; , \label{eq:1Y234}
\end{eqnarray}
which displays the considered symmetry.

The integrals involving higher tensors are treated in exactly
the same way. For the second-rank tensor we
use~(\ref{eq:compmn}). The reduction of $Y_{2 3 4}^{1 1}$ leads
to a formula involving $Y_{2 3 4}^{1 5}$. For this an expression
analogous to~(\ref{eq:1y234b}) is obtained. The symmetry
properties of the integral become manifest after using a
relation involving $T$-integrals.
The expressions for $Y_{2 3 4}^{1 1}$ and $Y_{2 3 4}^{1 5}$ are
listed in the appendix.

With the technique for the tensor decomposition of two-loop
tensor integrals described above all $Y$-integrals can be
reduced to scalar integrals of simpler structure
where no integration momentum
appears in the numerator, i.e.~to $T$-integrals.
The class of $T$-integrals therefore suffices to express all
possible two-loop self-energies.

This is true in every renormalizable gauge. For
practical purposes we briefly examine the situation in the
't Hooft-Feynman gauge. As explained above the most complicated
$Y$-integrals have denominators of the form shown in
fig.~\ref{fig:top2}. The possible numerators can be
obtained by power counting. For $\xi_i = 1$ the gauge boson
propagators~(\ref{eq:gabopr}) do not contribute powers of
momenta to the numerator of the Feynman integrals. Counting the
powers arising from the other Feynman rules and from the
decompositions~(\ref{eq:stens1a}),~(\ref{eq:stens2a}),
{}~(\ref{eq:stens3a1})
and~(\ref{eq:stens3a2}) reveals that the relevant integrals are
$Y_{2 3 4}^{1}$, $Y_{2 3 4 5}^{1}$, $Y_{2 3 4 5}^{1 1}$, $Y_{2 3
4 5 5}^{1}$ and $Y_{2 3 4 5 5}^{1 1}$. Explicit formulae for
these integrals are given in this section and in the appendix
where we also list some reduction formulae needed for general
values of $\xi_i$.

By using the methods outlined above the task of evaluating each
Feynman amplitude corresponding to a two-loop self-energy is
reduced to the calculation of a relatively
small number of scalar integrals
possessing a simpler structure. This is very desirable
since
a direct evaluation of the
tensor integrals appearing in the Feynman amplitude
by means of Feynman parameters in general yields
a large number of different integrals which all have to be treated
separately.
In contrast to
that the $T$-integrals are well suited for studying their
analytical properties as was already done by a number of
authors.

In~\cite{raibas,basdav,broad1,broad2}, for example, results in terms of
polylogarithmic functions were obtained for special cases of
masses and momenta.
In more general cases this class of functions is not
sufficient.
As was shown in~\cite{raidip}, the general massive
two-loop integrals cannot be
expressed in terms of Nielsen polylogarithms with arguments
being algebraic functions of the external variables.
For the ``master integral'' (fig.~\ref{fig:top1}) an integral
representation suitable for numerical evaluation was
derived in~\cite{krei}.
\chapter{Relations between the scalar integrals}

The results obtained for the two-loop self-energies via the steps
described in the preceding sections consist of a Lorentz
tensor specified in~(\ref{eq:stens1}),~(\ref{eq:stens2})
and~(\ref{eq:stens3}), respectively, and a scalar part being a
sum of $T$-integrals multiplied by rational functions in the squared
momentum $p^2$, the space-time dimension $D$, the particle
masses $m_j$ and the gauge parameters $\xi_i$. The algebraic
structure of the result is convenient for studying the
dependence on these variables, either by inserting the exact
results for the integrals, if these are known, or suitable
approximations, e.g.~in the asymptotic or threshold regime of
$p^2$ or in the limit $D \rightarrow 4$.

The $T$-integrals occurring in the result, which
we denote as $T^1, T^2, \ldots , T^n$ for the moment,
are in general not algebraically
independent of each other, i.e.~there exist relations
\begin{equation}
c^1 T^1 + c^2 T^2 + \ldots + c^n T^n  = 0 \;  .
\label{eq:relTn}
\end{equation}
The coefficients $c^1, \ldots , c^n$ are polynomials in $p^2, D,
m_j$ and the gauge parameters.
The results can be made very transparent by using these relations
to eliminate as many integrals as possible.
For example, if one considers a gauge invariant set of amplitudes
this property directly manifests itself in the
algebraic result provided that the set of integrals used for
expressing the result is sufficiently small.
In this case all terms depending on the gauge parameters $\xi_i$
disappear from the result. This means that the prefactors of all
gauge dependent basic integrals algebraically add up to zero while all
$\xi_i$-dependent terms multiplying a gauge independent
basic integral exactly cancel each other.
In this way the gauge invariance is seen to hold exactly,
i.e.~by purely algebraical means.
No explicit analytical or numerical expressions of the standard
integrals are needed for this consideration.

In the same way
as the gauge invariance
Slavnov-Taylor identities or any other relation involving two-loop
self-energies can be checked exactly. This is very useful for
performing consistency checks on the result.

The question arises whether a sufficiently small
basis of integrals can actually be found.
We argue that after
invoking the symmetry properties of the two-loop integrals and
eliminating all $Y$-integrals by means of tensor decompositions
only few additional relations between the scalar integrals
are needed. In sect.~\ref{ST} we will
verify Slavnov-Taylor identities valid for the self-energies of
the photon and the $Z$-boson by adding up the results of several
thousand Feynman amplitudes. For this application only the
relations for the integrals $T_{2 3 4 5'}(m_2^2, m_3^2, m_4^2)$
and $T_{2 3 4 5' 5'}(m_2^2, m_3^2, m_4^2)$ described below are used.

The first type of relations we want to consider
involves integrals in which at least one
propagator is massless.
They are obtained indirectly by using symmetry arguments,
momentum conservation and properties of the tensor
decomposition.
In this way  we derive a relation
between the integral $T_{2 3 4 5'}(m_2^2, m_3^2, m_4^2)$ and the
integrals of the same type where the masses are permuted.
It is a consequence of momentum conservation at the
four-vertices of the topology $\Delta _{2 3 4}$. Starting from
\begin{equation}
 \langle \langle \frac{ p \cdot (k_2 + k_3 - k_4 - p)}{ \Bigl[
 k_2^2 - m_2^2 \Bigr] \Bigl[ k_3^2 - m_3^2 \Bigr] \Bigl[ k_4^2 -
 m_4^2 \Bigr] } \rangle \rangle = 0 \; ,
\end{equation}
which obviously follows from~(\ref{eq:momenta}), using the
symmetries of the resulting integrals with respect to
permutations of $k_2, k_3$ and $k_4$ and performing tensor
decompositions for these leads to the relation
\begin{eqnarray}
&& \Bigl\{ (m_2 ^2 - m_3 ^2) (m_4 ^2 - p^2)
T_{2 3 4 5'}(m_2^2,m_3^2,m_4^2) +
\mbox{cycl.} \Bigr\}   \no
&& + \Bigl\{ (m_2 ^2 - m_3 ^2) T_{2 3 5'}(m_2^2,m_3^2)
 + \mbox{cycl.} \Bigr\} \no
&& - \Bigl\{ \Bigl[A_0(m_2^2) - A_0(m_3^2)\Bigr]
(m_4 ^2 - p^2) B_0(p^2;m_4^2,0) +
\mbox{cycl.} \Bigr\} = 0 \; .
 \label{eq:Trel1}
\end{eqnarray}
Here ``cycl.'' denotes cyclic permutation of the masses.
With~(\ref{eq:Trel1}) one permutation of the $T_{2 3 4 5'}$-integral can
always be eliminated. This relation was used in the
last section to obtain~(\ref{eq:1Y234}). If two masses are
equal,~(\ref{eq:Trel1}) becomes trivial.

Two similar relations hold for the permutations of the integral
$T_{2 3 4 5' 5'}$. Using these, it is possible to obtain one standard
permutation for every integral of this type.
In contrast to~(\ref{eq:Trel1}) this relation remains nontrivial
if two of the three masses are equal.
It gives an expression for the
integral $T_{2 3 4 5' 5'}(m^2, M^2, M^2)$ in terms of integrals with
fewer propagators only.
Therefore it can completely be replaced by
integrals of a simpler type.
We list this relation in the appendix. It takes a particularly
simple form
in the special case where $M^2 = 0$:
\begin{eqnarray}
 T_{2 3' 4' 5' 5'}(m^2) &=& \frac{1}{D m^2 p^2} \Biggl\{
 A_0(m^2) \Bigl[ (D - 4) B_0(p^2; 0, 0) +
 D p^2 B'_0(p^2; 0, 0) \Bigr]
 \no
 &&- (D - 2)^2 T_{2 3' 5'}(m^2)
 + (3 D - 8) T_{2 3' 4'}(m^2) \no
 &&- (D - 4) (m^2 + p^2) T_{2 3' 4' 5'}(m^2) \Biggr\} \; .
 \label{eq:trelpr}
\end{eqnarray}
The one-loop integral $B'_0$ is defined as
\begin{equation}
 B'_0(p^2;m_1^2,m_2^2) = \frac{ \partial}{\partial
 (m_1^2)}B_0(p^2;m_1^2,m_2^2) \; .
\end{equation}
Relations similar to those valid for $T_{2 3 4 5'}$ and
$T_{2 3 4 5' 5'}$
can be obtained for integrals with more massless propagators,
e.g.~$T_{2 3 4 5' 5' 5'}$ and $T_{2 3 4 5' 5' 5' 5'}$.

A second class of relations can be derived via the well known
method of integration by parts. It yields relations for
integrals where at least one propagator appears with a power
higher than one. A particularly interesting example is
the formula for the integral
corresponding to topology 6 in fig.~\ref{fig:top},
i.e.~$T_{1 1 3 4}(m_1^2, m_1^2, m_3^2, m_4^2)$. Starting from
\begin{equation}
 D \, T_{1 3 4} = \langle \langle \Delta _{1 3 4} \frac{ \partial
 k_1^{\mu}}{\partial k_1^{\mu}} \rangle \rangle = - \langle
 \langle k_1^{\mu} \frac{ \partial}{\partial k_1^{\mu}} \Delta
 _{1 3 4} \rangle \rangle
\end{equation}
and using the symmetry of the integral $T_{1 3 4}$ with respect
to permutation of the masses $m_1^2$ and $m_3^2$ leads to
\begin{eqnarray}
 T_{1 1 3 4}&=& \frac{- 1}{\lambda (m_1^2, m_3^2, m_4^2)}
 \Biggl\{ (3-D) (m_1^2 - m_3^2 - m_4^2) T_{1 3 4} -
 (1 - D/2)
 \Bigl[ A_0(m_1^2) \Bigl( A_0(m_3^2) \no
 &&+ A_0(m_4^2) \Bigr) - 2
 A_0(m_3^2) A_0(m_4^2) \Bigr] - (m_3^2 - m_4^2) \Bigl( A_0(m_3^2)
  -
 A_0(m_4^2) \Bigr) B_0(0; m_1^2, m_1^2) \Biggr\}
 \; , \no
 && \label{eq:T1134}
\end{eqnarray}
where
\begin{equation}
\lambda (m_1^2, m_3^2, m_4^2) =  m_1^4 + m_3^4 + m_4^4 -
 2 (m_1^2 m_3^2 + m_1^2 m_4^2 + m_3^2 m_4^2) \; .
\end{equation}
The integral $T_{1 1 3 4}$ can therefore
be expressed through $T_{1 3 4}$ and
products of one-loop integrals.

Finally it should be noted that some
integrals vanish trivially in the framework of
dimensional regularization, e.g.
\begin{equation}
 T_{1 1} = T_{1 2} = T_{1 1 2} = 0 \; .
\end{equation}
\chapter{Computer-algebraic realization}

The procedure we described for the algebraic calculation of two-loop
self-energies is algorithmic
and we implemented it
into a computer-algebra program which we called \two \cite{two}. It is
written in \ma and linked to the packages \fea \cite{fea} and
\fec \cite{fec}. \fea creates the Feynman amplitudes and draws
the Feynman graphs. \fec is used here for
the contraction
of Lorentz indices and the evaluation of Dirac-traces.
In the calculations performed in this paper an anticommuting
$ \gamma _5$ in $D$ dimensions was used.

The three packages run fully
automatically. This high degree of automation is very convenient for
performing calculations which involve a large number of
diagrams.
\chapter{Slavnov-Taylor identities for two-loop self-energies}
\label{ST}

In order to demonstrate the abilities of our algorithm for
doing large calculations
we explicitly verify some
Slavnov-Taylor identities valid for two-loop self-energies.
In this section we use the 't Hooft-Feynman gauge.

In~\cite{ren} exact relations for the gauge boson propagators
$\Delta_{\mu \nu }^{ \alpha \beta }$, the gauge boson Higgs
mixing propagators $\Delta_{\mu}^{ \alpha i}$ and the unphysical
Higgs propagators $\Delta^{i j}$
are listed
\begin{equation}
 \begin{array}{lclllll}
p^{\mu} p^{\nu} \Delta^{\gamma \gamma}_{\mu \nu} (p)& & & &  & =
& - i \\
p^{\mu} p^{\nu} \Delta^{Z Z}_{\mu \nu} (p) &-& 2i M_{Z} p^{\mu}
\Delta^{Z \chi}_{\mu} (p) &+& M_{Z}^2 \Delta^{\chi \chi}
(p) &
= & - i \\
p^{\mu} p^{\nu} \Delta^{\gamma Z}_{\mu \nu} (p) &-& i
M_{Z} p^{\mu}
\Delta^{\gamma \chi}_{\mu} (p) & & & = & 0 \\
 p^{\mu} p^{\nu} \Delta^{W W}_{\mu \nu} (p) &+& 2 M_{W} p^{\mu}
 \Delta^{W \varphi}_{\mu} (p) &+& M_{W}^2 \Delta^{\varphi \varphi} (p) &
 = & -
 i \; ,
\end{array}
\end{equation}
where $\chi$ and $\varphi$ are the neutral and charged unphysical
Higgs fields, respectively.
Writing the propagators as the inverse of the truncated one-particle
irreducible two-point functions as specified in
sect.~\ref{classif}
and expanding up to
second order leads to the second-order relations
\begin{equation}
\begin{array}{lcl}
p^2 \Bigl[ \Sigma ^{ \gamma \gamma ,(2)}_{L} \; \: - \Bigl(
\Sigma ^{ \gamma
\chi
,(1)} \Bigr)^2 \; \; \; \;  \, \Bigr] & = & 0  \\
p^2 \Bigl[ \Sigma ^{Z Z ,(2)}_{L} \; - 2 i  M_{Z} \Sigma ^{Z
\chi
,(2)}
\: \Bigr] -  M_{Z}^2 \Sigma ^{ \chi \chi ,(2)} \: - p^2 \Bigl(
\Sigma ^{Z
\chi ,(1)} \: \Bigr)^2  + \Sigma ^{Z Z ,(1)}_{L} \Sigma ^{ \chi
\chi
,(1)} & = & 0 \\
p^2 \Bigl[ \Sigma ^{ \gamma Z ,(2)}_{L} \; \, - i M_{Z} \Sigma
^{
\gamma
\chi ,(2)} \; \; \; \Bigr]  - p^2 \Sigma ^{ \gamma \chi ,(1)}
\Sigma ^{Z
\chi ,(1)} \: + i  M_{Z} \Sigma ^{ \gamma \chi ,(1)}
\Sigma ^{ \chi \chi ,(1)} & = & 0 \\
p^2 \Bigl[ \Sigma ^{W W ,(2)}_{L} + 2 M_{W} \Sigma ^{W \varphi ,(2)}
\Bigr] -
M_{W}^2 \Sigma ^{ \varphi \varphi ,(2)} - p^2 \Bigl( \Sigma ^{W \varphi
,(1)} \Bigr)^2 + \Sigma ^{W W ,(1)}_{L} \Sigma ^{ \varphi \varphi ,(1)}
& = & 0 \; . \label{eq:self2}
\end{array}
\end{equation}
The superscripts $^{(n)}$ indicate the two-loop and one-loop
self-energies, respectively.
Similar relations can be obtained for the one-loop
self-energies~\cite{ren}.
As always, it is understood that the tadpole contributions
are part of the self-energies (see fig.~\ref{fig:top}).

The first relation in~(\ref{eq:self2})
indicates that unlike in pure QED the longitudinal part of the two-loop
photon self-energy does not vanish.
We used \two to calculate explicitly
every Feynman amplitude contributing
to the left hand side of this relation.
We did this in the full electroweak SM with the only restriction that
the fermions are limited to one doublet, which we chose to
be the $u$- and $d$-quark. The calculations for all other
fermions can be performed analogously.

We had to consider $1024$ graphs for the two-loop photon self-energy,
$162$ of which contain the $u$- and $d$-quarks, the remaining
ones are purely bosonic. The most involved diagrams are of
course those stemming from the generic two-loop topologies.
These are $418$ diagrams belonging to the first 6 topologies in
fig.~\ref{fig:top}. The remaining ones lead to integrals
expressible solely as products of one-loop integrals.
Altogether $13$ of the $20$ possible
topologies
are realized for the photon self-energy.

The one-loop $( \gamma \chi )$-mixing self-energy vanishes for
fermionic graphs, while $4$ purely bosonic graphs
contribute.
Summing up the results of all amplitudes gives exactly zero,
i.e.~the prefactor of every standard integral vanishes
separately.

We also explicitly verified the second relation stated
in~(\ref{eq:self2}). Here, of course, a lot more Feynman
diagrams have to be calculated. Again we restricted the fermions
to $u$- and $d$-quark. This gives $1552$ diagrams for $\Sigma ^{Z Z
,(2)}_{L}$, $1155$ for $\Sigma ^{Z \chi ,(2)}$ and $1142$ for
$\Sigma ^{ \chi \chi ,(2)}$. From these, $1787$ diagrams arise
from the $8$ generic two-loop topologies.

For the self-energies of the $Z$-boson and the unphysical
Higgs field $\chi$ all $20$ topologies contribute which are maximally
possible for two-loop self-energies.
The self-energies in one-loop order give rise to $65$
diagrams. Adding those to the $3849$ two-loop graphs by using
the coefficients
specified in~(\ref{eq:self2}) gives exactly zero.
\chapter{Results for the electroweak Standard Model and
discussion}

As an application of the techniques described above
we treat the
gauge boson self-energies in the electroweak SM.
We begin with the special case of the
photon self-energy in pure QED. The three contributing diagrams
are shown in fig.~\ref{fig:qedgraphs}.
\begin{figure}
\begin{picture}(350,80)
\end{picture}
\caption{The diagrams contributing to the two-loop photon self-energy
in QED.}
\label{fig:qedgraphs}
\end{figure}

We decompose the photon self-energy according
to~(\ref{eq:stens1}). The longitudinal part vanishes due to the
well-known Ward-identity. We checked this by explicit
calculation. The result for the transverse part
reads
\begin{eqnarray}
 \Sigma ^{ \gamma \gamma ,(2)}_{T, QED}(p^2) &=& \frac{e^4}{128
 \pi^4 (D - 1)} \Bigg\{ \Bigl[(2 - D)^3 /m_e^2
 \Bigr] A^2_0(m_e^2) -
  \Bigl[(2 - D)^2 (m_e^2 - p^2)/m_e^2 \Bigr]  \no
 && \times A_0(m_e^2) B_0(p^2; 0, m_e^2) -
  \Bigl[(2 - D)
  (4 m_e^2 + 2 p^2 - D p^2)/m_e^2 \Bigr] A_0(m_e^2) \no
 && \times B_0(p^2; m_e^2, m_e^2)
 - (8 m_e^2 - 14 p^2 + 9 D p^2 - D^2 p^2) B^2_0(p^2;
 m_e^2, m_e^2)\no
 &&-  (2 - D)
 \Big[ 2 (4 m_e^2 - 2 p^2 + D p^2) A_0(m_e^2)
 B'_0(p^2; m_e^2, m_e^2) + (2 - D) T_{13'4'} \no
 && - 2 (1 - D) T_{134'}
  - (6 - D) T_{234'}
  + (2 - D) (m_e^2 - p^2) T_{1'234'} \Big]\no
  &&+ 8 (4 m_e^2 - 2 p^2 + D p^2)
 \Big[ T_{1234'} +  m_e^2 T_{11234'} \Big] \no
 && + 2 (2 m_e^2 - p^2) (4 m_e^2 - 2 p^2 + D
 p^2) T_{123'45} \Bigg\} \; . \label{eq:photqed}
\end{eqnarray}
As above a prime at the subindices of the $T$-integrals denotes
that the corresponding propagator has mass zero. All other
propagators carry the electron mass $m_e$.

The calculation was carried out in an arbitrary $R_{\xi}$-gauge.
The absence of the gauge parameter $\xi_{ \gamma }$ indicates
that the result is gauge invariant as it has to be. The
two-loop integrals appearing in~(\ref{eq:photqed}) are expressible
in terms of polylogarithmic functions,
i.e.~logarithms, di- and trilogarithms. They were
studied for example in~\cite{broad1,broad2}.
The one-loop integrals arising from the calculation of
irreducible two-loop contributions are in general needed up to
${\cal O}(D - 4)$. Results for these can be found
in~\cite{raibas,oneloop}.

Calculations of the two-loop photon self-energy in QED were
performed in~\cite{QED}. Similar calculations
focusing on
the ${\cal O}$($ \alpha \alpha _s$) corrections due to the
gluon exchange in
a quark loop of gauge boson self-energies
were done in~\cite{QCD}.

In passing we note that in contrast to the photon self-energy
the two-loop electron
self-energy in QED is not gauge invariant. The explicit algebraic
result will be presented elsewhere. It involves the integrals
$T_{234}, T_{1'234}$ and $T_{1'2345'}$ which cannot be reduced
to polylogarithmic functions. The problem of evaluating
$T_{1'2345'}$ was addressed in~\cite{broad1}.

In the electroweak SM results for two-loop
gauge boson self-energies
restricted to the region $p^2
= 0$ and asymptotic values of the top and Higgs mass,
respectively, were worked out in~\cite{vdBH,barb2,vdBV}.
A calculation allowing for general values of the invariant
momentum $p^2$ and the masses of the gauge bosons, the top quark
and the Higgs field has to deal with an enormous number of Feynman
diagrams having in general a very complicated structure.
With the methods outlined above we can express these in terms of
a small number of standard integrals, the result being valid for
general values of all parameters involved.

This was explicitly carried out in sect.~\ref{ST} for all graphs
contributing to the longitudinal parts of the photon and
$Z$-boson self-energies.
The results for the complete transverse parts are of course
rather lengthy and we do not present them here. Instead we focus
on the light fermion contributions, a class of diagrams
particularly well suited for separate treatment which is
expected to yield a substantial contribution to the complete
result.

With light fermion contributions we mean all diagrams containing
any fermion other than the top-quark. These fermion masses are
negligible in comparison to the boson masses ($m_Z,
m_W, m_H$) and the invariant momentum $p^2$, if  $p^2$ is
sufficiently large. All $T$-integrals resulting from
these graphs therefore contain one massless fermion loop. This
feature allows an analytic evaluation resulting in
polylogarithmic functions. We have explicitly calculated all
integrals needed for the two-loop gauge boson self-energies,
i.e.~all integrals appearing in the algebraic results stated
below. The results will be presented in a related
paper~\cite{raibas}.

In order to study the structure of the results we find it
sufficient to restrict the fermions to one doublet in the same
way as in sect.~\ref{ST}. This yields 27 diagrams contributing
to the two-loop photon self-energy, 32 graphs for the $Z$
self-energy and 45 diagrams for the self-energy of the $W^{\pm}$.

As shown in fig.~\ref{fig:phot} it is convenient to subdivide
the photon self-energy into several sets of graphs.
The graphs depicted in fig.~\ref{fig:phot}a are of QED-type
having a photon exchanged in a quark loop. As indicated in the
picture the same set of graphs also exists for the $u$-quark.
The results are in exact analogy to~(\ref{eq:photqed}) and we do
not write them out explicitly.
As a consequence of~(\ref{eq:photqed}) the QED-type graphs
form gauge invariant subsets of the photon self-energy.

In the graphs shown in fig.~\ref{fig:phot}b a $Z$-boson is
exchanged instead of the $ \gamma $. We give the result
neglecting the mass of the $d$-quark:
\begin{eqnarray}
 \Sigma ^{ \gamma \gamma ,(2)}_{T, Z}(p^2) &=& \frac{e^4 (2 - D)
 (9 - 12 s_W^2 + 8 s_W ^4)}{(4 \pi)^4 (1 - D) D 108 c_W^2
 s_W^2} \Bigg\{ 2 (4 - 6 D + D^2) A_0(m_Z^2) B_0(p^2; 0, 0)  \no
 && - D
 (2 m_Z ^2 + 7 p^2 - D p^2) B^2_0(p^2; 0, 0)
 + 2 (4 - 6 D + D^2) T_{13'4'}(m_Z^2) \no
 && + 2 (8 - 4 D + D^2) T_{23'4'}(m_Z^2) - 2 (4 - 8 D + D^2)
 (m_Z^2 + p^2) T_{1'2'34'}(m_Z^2) \no
 && - D (2 m_Z^4 + 8 m_Z^2 p^2 - D m_Z^2 p^2 + 2 p^4)
 T_{1'2'34'5'}(m_Z^2) \Bigg\} \; . \label{eq:photZ}
\end{eqnarray}
The subscript $Z$ refers to the $Z$-boson exchange. We used
the abbreviations
\begin{equation}
 c_W^2 = \frac{m_W^2}{m_Z^2} \; , \; \; s_W^2 = 1 -
 \frac{m_W^2}{m_Z^2} \; .
\end{equation}
It can be seen from~(\ref{eq:photZ}) that for non-exceptional
values of $p^2$ no mass singularities are induced by setting the
fermion mass to zero (see~\cite{kino}).

There is no gauge dependence left in~(\ref{eq:photZ}).
Therefore all contributions
from ``neutral currents'', i.e.~$ \gamma $ and $Z$ exchange,
are gauge invariant.

\begin{figure}
\begin{picture}(453,500)
\put(220,480){$(a)$}
\put(272,524){,}
\put(352,524){$d \leftrightarrow u$}
\put(220,335){$(b)$}
\put(272,380){,}
\put(352,380){$d \leftrightarrow u$}
\put(220,10){$(c)$}
\end{picture}
\caption{The light fermion contributions to the two-loop photon
self-energy}
\label{fig:phot}
\end{figure}

The ``charged currents'' mediated by $W^{\pm}$ and $\varphi ^{\pm}$
give rise to the 15 diagrams shown in fig.~\ref{fig:phot}c. The
result for the transverse part reads
\begin{eqnarray}
 \Sigma ^{ \gamma \gamma ,(2)}_{T, W^{\pm}}(p^2) &=& \frac{e^4 (2 -
 D)}{(4 \pi)^4 (1 - D)^2 12 m_W^4 s_W^2} \Bigg\{ f(p^2, m_W^2,
 D) + 9 \, g(\xi_W, p^2, m_W^2, D) \no
 &&+ F(p^2, m_W^2, D) + 9 m_W^2 (2 m_W^2 - p^2) G(\xi_W, p^2,
 m_W^2, D) \Bigg\} \label{eq:photW} \; ,
\end{eqnarray}
where in $f(p^2, m_W^2, D)$ and $g(\xi_W, p^2, m_W^2, D)$
we collected
the contributions involving only one-loop integrals
\begin{eqnarray}
f(p^2, m_W^2, D)&=&
 2 m_W^2 \Big[ \Big( \frac{40}{D} - 91  + 70 D - 10 D^2 \Big) m_W^2
 - 18 (2 - D) p^2\Big] A_0(m_W^2) B_0(p^2; 0, 0) \no
 &&+ 8 (1 - D) m_W^4 \Big[ 2 m_W^2 + (7 - D) p^2 \Big]
 B_0^2(p^2; 0, 0)
 - 9 (2 m_W^2 - p^2) \no
 && \times \Big[ 4 (1 - D) m_W^4  +
  4 (3 - 2 D) m_W^2 p^2  - p^4 \Big] B_0(p^2; 0, 0) B_0(p^2; m_W^2,
  m_W^2) , \no
  &&\label{eq:Gf}
\end{eqnarray}
\begin{eqnarray}
g(\xi_W, p^2, m_W^2, D)&=&
 - \frac{2 m_W^2}{\xi_{W}} (m_W^2 - p^2) A_0(m_W^2) B_0(p^2; 0, 0)
 + 2 m_W^2 \Big[ \frac{1}{\xi_W} (m_W^2 - p^2) - m_W^2 \no
 &&+ 2 (2 - D) p^2 \Big] A_0(m_W^2/ \xi_W) B_0(p^2; 0, 0)
  - 2 (m_W^2 - p^2) \Big[
   m_W^4 \Big( \frac{1}{\xi_{W}} - 1 \Big)^2 \no
 && - 2 \Big( \frac{1}{\xi_{W}} + 3 - 2 D \Big) m_W^2 p^2
  + p^4 \Big] B_0(p^2; 0, 0)
 B_0(p^2; m_W^2, m_W^2/ \xi_W) \no
 &&+ p^4 \Big(4 \frac{m_W^2}{\xi_{W}}
  - p^2 \Big) B_0(p^2; 0, 0) B_0(p^2; m_W^2/ \xi_W, m_W^2/
 \xi_W) \label{eq:Gg}  ,
\end{eqnarray}
and $F(p^2, m_W^2, D)$ and $G(\xi_W, p^2, m_W^2, D)$ contain the
generic two-loop contributions
\begin{eqnarray}
F(p^2, m_W^2, D)&=& m_W^2 \Bigg[ \Big[ 2 (
 \frac{40}{D} - 118 + 187 D - 100 D^2 + 18 D^3 ) m_W^2
 - 9 (7 - 4 D) (3 - D) p^2 \Big] \no
 && \times T_{13'4'}(m_W^2)
 + \Big[4 ( \frac{40}{D} - 96 + 61 D - 14 D^2 ) m_W^2
 + 9 p^2 \Big] T_{23'4'}(m_W^2) \no
 &&- 20 (1/D - 1) (4 - 8 D + D^2) m_W^2 (m_W^2 + p^2) T_{1'2'34'}(m_W^2)
 + 18 (4 m_W^2 \no
 &&- p^2) \Big[ 2 (1 - D) m_W^2 + (3 - 2 D) p^2
 \Big] T_{123'4'}(m_W^2, m_W^2) + 9 (4 m_W^2 - p^2) \no
 && \times \Big[ 4 (1 - D) m_W^4 + 4 (3 - 2 D) m_W^2 p^2 - p^4 \Big]
 T_{1123'4'}(m_W^2, m_W^2, m_W^2) \no
 &&+ 72 (1 - D) m_W^4 (m_W^2 + 2 p^2)
  T_{123'4'5'}(m_W^2, m_W^2) \no
 &&
 + 8 (1 - D) m_W^2 \Big[2 (m_W^4 +
 p^4) + (8 - D) m_W^2 p^2 \Big] T_{1'2'34'5'}(m_W^2) \Bigg]  ,
 \label{eq:GF}
\end{eqnarray}
\begin{eqnarray}
G(\xi_W,p^2, m_W^2, D) &=&
 \frac{1}{\xi_W} (D - 3) T_{13'4'}(m_W^2) +
 T_{23'4'}(m_W^2/ \xi_W) - 2 \Big[ \Big(\frac{1}{\xi_W}
 - 1\Big) m_W^2 \no
 &&+ (3 - 2 D) p^2 \Big]
 T_{123'4'}(m_W^2, m_W^2/ \xi_W) +
 \Big[m_W^4 \Big( \frac{1}{\xi_{W}} - 1 \Big)^2 \no
 && - 2 \Big( \frac{1}{\xi_{W}} + 3 - 2 D \Big) m_W^2 p^2
  + p^4 \Big]
 T_{1123'4'}(m_W^2, m_W^2, m_W^2/ \xi_W) . \label{eq:GG}
\end{eqnarray}
The functions $f(p^2, m_W^2, D)$ and $F(p^2, m_W^2, D)$ are
independent of the gauge parameter $\xi_W$.
The occurrence of $\xi_W$ in  $g(\xi_W, p^2, m_W^2, D)$ and
$G(\xi_W, p^2, m_W^2, D)$ indicates that
the contribution from the ``charged current'' graphs is gauge
dependent. This feature is expected since also in one-loop order
the graphs containing $W$-bosons give gauge dependent contributions to
the photon self-energy (see for example~\cite{Sirlin}).

As explained above the integrals occurring in~(\ref{eq:photW}) can
be solved analytically for general values of $\xi_W$.
Using the expressions given in~\cite{raibas} the dependence on
the variables $p^2, m_W^2, D$ and $\xi_W$ can be studied.

For all results contributing to the photon
self-energy we also calculated the longitudinal part.
It vanishes separately for each set of graphs specified above and
without restriction on the values of the gauge parameters.
We also studied the limit $p^2 \rightarrow 0$ for all results given
above and checked that they give zero for all values of the gauge
parameters and the dimension $D$.

Next we focus on the self-energy of the $Z$-boson. We first note
that there are contributions from 27 graphs corresponding to the
ones considered for the photon self-energy, i.e.~from the
diagrams shown in fig.~\ref{fig:phot} where the in- and outgoing
photon is substituted by a $Z$. In addition to these,
five graphs occur containing the Higgs-field, three of which are
tadpole graphs. These diagrams are depicted in fig.~\ref{fig:Z}.

\begin{figure}
\begin{picture}(453,80)
\end{picture}
\caption{Higgs-dependent contributions to the $Z$ self-energy}
\label{fig:Z}
\end{figure}

We begin with the diagrams which are of the same form as in the
case of the photon. The transverse parts of the ``neutral
current'' graphs are treated in precisely the same way as
described above.
The results have the same form and contain exactly the same
integrals as the corresponding ones for the photon self-energy.
Since we are only interested in the structure of the results
we do not list them here explicitly. As above, these graphs
yield gauge invariant contributions.

For the 15 ``charged current'' graphs we obtain:
\begin{eqnarray}
 \Sigma ^{Z Z,(2)}_{T, W^{\pm}}(p^2) &=& \frac{e^4 (2 -
 D)}{(4 \pi)^4 (1 - D)^2 12 m_W^2 m_Z^2 s_W^4} \Bigg\{
 \tilde{f}(p^2, m_W^2, m_Z^2, D) \no
 &&+ 9 \, g(\xi_W, p^2, m_W^2, D)
 + \tilde{F}(p^2, m_W^2, m_Z^2, D) \no
 &&+ 9 (p^2 - m_Z^2) (2 m_W^2 - m_Z^2 -
 p^2) \frac{m_W^2}{p^2} G(\xi_W,p^2, m_W^2, D)
 \Bigg\} \; .  \label{eq:zW}
\end{eqnarray}
The gauge dependent functions $g(\xi_W, p^2, m_W^2, D)$ and
$G(\xi_W,p^2, m_W^2, D)$ are the same as for the photon
self-energy. They were given in~(\ref{eq:Gg}) and~(\ref{eq:GG}),
respectively. The expressions for the gauge independent
functions $\tilde{f}(p^2, m_W^2, m_Z^2, D)$ and $\tilde{F}(p^2, m_W^2,
m_Z^2, D)$ read
\begin{eqnarray}
 \tilde{f}(p^2, m_W^2, m_Z^2, D) &=&
 2 \Big[ (1/D - 1) (4 - 6 D + D^2) \Big( 9 m_W^4 + (m_W^2 -
 m_Z^2)^2 \Big)
 + 9 m_W^2 \Big( m_W^2 \no
 &&- 2 (2 - D) p^2 \Big) \Big] A_0(m_W^2) B_0(p^2; 0, 0)
 + (1 - D) (4 m_W^2 - m_Z^2) (2 m_W^2 \no
 &&+ m_Z^2) \Big[2 m_W^2 + (7  - D) p^2
 \Big] B_0^2(p^2; 0, 0)
 - 9 (2 m_W^2 - p^2) \Big[ 4 (1 - D) m_W^4  \no
 &&+ 4 (3 - 2 D) m_W^2 p^2 - p^4 \Big]
 B_0(p^2; 0, 0) B_0(p^2; m_W^2,
 m_W^2) \label{eq:zf} \; ,
\end{eqnarray}
\begin{eqnarray}
\tilde{F}(p^2, m_W^2,  m_Z^2, D) &=&
  \Big[ 2 \Big(\frac{40}{D} - 118
 + 187 D - 100 D^2 + 18 D^3 \Big) m_W^4 - \Big( \frac{8}{D}
 + 7 + 5 D - 2 D^2 \no
 &&+ 9 (2 - D) \frac{m_W^2}{p^2} \Big)
 m_Z^2 ( 2 m_W^2 - m_Z^2 ) - 9 (7 - 4 D) (3 - D)
  m_W^2 p^2 \Big] \no
 && \times T_{13'4'}(m_W^2)
 + \Big[ 4 (\frac{40}{D} - 96  + 61 D - 14 D^2 )
 m_W^4 - 2 (1/D - 1) \no
 && \times (8 - 4 D + D^2) m_Z^2 (2 m_W^2 - m_Z^2 )
 + 9 m_W^2 p^2 \Big] T_{23'4'}(m_W^2) \no
 &&-  2 (1/D - 1) (4 - 8 D
 + D^2) (m_W^2 + p^2) \Big( 9 m_W^4 + (m_W^2 - m_Z^2)^2 \Big) \no
 && \times T_{1'2'34'}(m_W^2)
 + 18 m_W^2 (4 m_W^2 - p^2) \Big[ 2 (1 - D) m_W^2
 + (3 - 2 D) p^2 \Big] \no
 && \times T_{123'4'}(m_W^2, m_W^2)
 + 9 m_W^2 (4 m_W^2 - p^2) \Big[ 4 (1 - D) m_W^4 \no
 &&+ 4 (3 - 2 D) m_W^2 p^2 - p^4 \Big] T_{1123'4'}(m_W^2, m_W^2, m_W^2)
 \no
 &&+ 72 (1 - D) m_W^6 (m_W^2
 + 2 p^2) T_{123'4'5'}(m_W^2, m_W^2)
 + (1 - D) (4 m_W^2 \no
 &&- m_Z^2) (2 m_W^2 + m_Z^2) \Big[2 (m_W^4 + p^4)
 + (8 - D) m_W^2 p^2 \Big]
  T_{1'2'34'5'}(m_W^2) \; . \no
 && \label{eq:zF}
\end{eqnarray}
For $m_Z^2 = 0$ the functions $\tilde{f}(p^2,
m_W^2, m_Z^2, D)$ and $\tilde{F}(p^2, m_W^2, m_Z^2, D)$ coincide
with the corresponding ones used for the photon self-energy,
i.e.~$f(p^2, m_W^2, D)$ and $F(p^2, m_W^2, D)$.

The characteristic feature of the result~(\ref{eq:zW})
is that all gauge dependent
contributions involving two-loop integrals are proportional to
$(p^2 - m_Z^2)$. This can be understood either by noting that
$p^2 = m_Z^2$ is the mass shell condition or, in the framework
of the intrinsic pinch technique~\cite{pinch},
by observing that this factor is necessary to
cancel a lowest-order $Z$-propagator.

\begin{figure}
\begin{picture}(453,630)
\put(220,100){$(a)$}
\put(220,0){$(b)$}
\end{picture}
\caption{The light fermion contributions to the $W$ self-energy}
\label{fig:W}
\end{figure}

In contrast to the photon case the light fermion contributions
to the $Z$ self-energy depend on the Higgs mass. Whereas the
contributions from the three tadpole graphs are compensated by
the usual choice of the renormalization constant, the first two
graphs in fig.~\ref{fig:Z} give a $p^2$-dependent contribution
to the $Z$ self-energy. We obtain for the first graph
\begin{eqnarray}
 \Sigma ^{Z Z,(2)}_{T, H}(p^2) &=& \frac{e^4 (2 -
 D) (9 - 12 s_W^2 + 8 s_W^4)}{(4 \pi)^4 (1 - D)^2 48 p^2 c_W^4 s_W^4}
 \Bigg\{ \Big[ (3 - D) (p^2 - m_H^2) + (2 - D) m_Z^2 \Big]
 T_{13'4'}(m_Z^2) \no
 &&+ m_Z^2 T_{23'4'}(m_H^2) - 2 m_Z^2 \Big[ m_H^2 -
 m_Z^2 + (3 - 2 D) p^2 \Big] T_{123'4'}(m_Z^2, m_H^2) \no
 &&+ m_Z^2 \Big[
 (m_H^2 - m_Z^2)^2 - 2 p^2 \Big( m_H^2 + (3 - 2 D) m_Z^2 \Big) +
 p^4 \Big] T_{1123'4'}(m_Z^2, m_Z^2, m_H^2) \Bigg\} , \no
 && \label{eq:zH}
\end{eqnarray}
which is obviously gauge invariant. The subscript $H$ indicates
the dependence on the Higgs mass which appears in the
scalar integrals and as coefficients $m_H^2$
and $m_H^4$, respectively.
The result for the second graph in fig.~\ref{fig:Z} is
derived from~(\ref{eq:zH}) by using the
appropriate coupling of the $u$-quark.

The result for the three tadpole graphs reads
\begin{eqnarray}
 \Sigma ^{Z Z,(2)}_{T, \, tad} &=& \frac{e^4 (2 -
 D)^2 m_Z^2}{(4 \pi)^4 12 m_H^2 s_W^4}
 \Bigg\{18 T_{13'4'}(m_W^2) + \Big[ 11 \frac{(m_W^2 - m_Z^2)^2}{m_W^4}
 + 9 \Big]
 T_{13'4'}(m_Z^2) \Bigg\} , \label{eq:ztad}
\end{eqnarray}
showing that they are gauge invariant. This is also
true for each of the graphs separately.

In the study of the $ \gamma Z$-mixing self-energy $\Sigma
^{\gamma Z,(2)}(p^2)$ no new
features appear. The diagrams are of the same type
as for the photon, i.e.~we have contributions from
the 27 graphs shown in fig.~\ref{fig:phot}
with one of the external photons substituted by a
$Z$-boson. Since the results have precisely the same structure as in
the case of the photon we do not list them in detail.

The graphs relevant for the $W^{\pm}$ self-energy are shown in
fig.~\ref{fig:W}.
As can be seen in fig.~\ref{fig:W}a there is no natural
way to make a subdivision into ``neutral'' and ``charged
current'' graphs.
Like in the case of the $Z$ self-energy we get
contributions from diagrams
depending on the Higgs mass. These are shown in
fig.~\ref{fig:W}b.

We first consider the 41 graphs listed in fig.~\ref{fig:W}a.
The result in 't Hooft-Feynman gauge, i.e.~$\xi_{ \gamma } =
\xi_{Z} = \xi_{W} = 1$, is given in the appendix.
Inspection of the integrals appearing in~(\ref{eq:w}) shows
that
the only integral type not present in the results of the
neutral gauge bosons is $T_{123'4'5'}(m_Z^2, m_W^2)$. Like the
integrals considered above it can be solved analytically. The
result is given in~\cite{raibas}.

The result for general values of the gauge parameters is very
lengthy and will not be presented here.
It depends on all three gauge parameters
{}~$\xi_{ \gamma }, \xi_{Z}$ and $\xi_{W}$. Therefore
in this case also the neutral
gauge bosons yield a gauge dependent contribution. In analogy to
the $Z$ self-energy all gauge dependent terms involving the
basic two-loop integrals are proportional to $(p^2 - m_W^2)$.
The integrals appearing in this result are the same as in the
case of the 't Hooft-Feynman gauge (see~(\ref{eq:w})).

The results for the Higgs-dependent graphs in fig.~\ref{fig:W}b
are of the same form as those obtained for the $Z$ self-energy
in~(\ref{eq:zH}) and~(\ref{eq:ztad}),
where in~(\ref{eq:zH}) $m_Z$ has to be substituted by $m_W$.
\chapter{Conclusion}

Radiative corrections are necessary for the comparison
between theory and precision measurements.
Whereas the methods for performing one-loop calculations
in massive gauge theories are well established, so far no
complete treatment of irreducible two-loop corrections
has been achieved. In the SM these corrections are
generally expected to be small, but a detailed analysis is
necessary to determine their actual value.

In this paper we presented a technique for reducing two-loop
self-energies to standard scalar integrals
in general massive gauge theories.
It makes use of the tensor structure of the two-loop integrals,
the symmetries of the scalar integrals and certain integral
relations. The results are valid for all values of the invariant
momentum $p^2$, the particle masses, the space-time dimension
$D$ and the gauge parameters.
Features like gauge invariance or transversality
are displayed directly at the algebraic level. We explicitly verified
Slavnov-Taylor identities valid for the neutral gauge boson
system of the SM by calculating several thousand Feynman diagrams.

As an application we treated the light fermion contributions
to the gauge boson self-energies in the SM.
For the self-energies of the neutral gauge bosons the contributions
associated with the exchange of $\gamma$ and $Z$ in the fermion
loop are gauge invariant. The $W$-exchange yields gauge dependent
contributions.  At the pole, i.e.~at $p^2 = m_Z^2$ for the $Z$
self-energy, the gauge dependence of the generic two-loop
contributions vanishes. The photon self-energy gives zero for
$p^2 = 0$ as required by the Ward identity.
The $Z$ self-energy acquires a Higgs mass dependent contribution
even in the light fermion case. We showed that it is gauge invariant.

The $W$ self-energy receives gauge dependent contributions from
all intermediate vector bosons. As in the case of the $Z$
self-energy the generic two-loop contributions are gauge invariant
at the pole, i.e.~for $p^2 = m_W^2$.
The Higgs-dependent graphs form a gauge invariant subset.

All standard scalar integrals needed for the treatment of the
light fermion contributions to the gauge boson self-energies
can be solved analytically. We performed these calculations
and obtained expressions in terms of polylogarithmic functions.
The results will be presented in a related paper~\cite{raibas}.

\vspace{1.0 cm}

We would like to thank F.A.~Berends, W.L.~van Neerven and
J.B.~Tausk for fruitful discussions, R.~Mertig for his
contributions to the program \two and for providing \fec
and
J.~K\"ublbeck and H.~Eck for help concerning \fea and its use
to create the pictures included in this paper.
\appendix

\chapter{Result for the $W$ self-energy}

The contribution from the 41 graphs listed in fig.~\ref{fig:W}a
to the two-loop self-energy of the $W$-boson is given by
\begin{eqnarray}
 \lefteqn{
 \Sigma ^{W W,(2)}_{T}(p^2) \Big|_{\xi_{ \gamma } = \xi_{Z} =
 \xi_{W} = 1} = \frac{e^4 (2 - D)}{(4 \pi)^4
 (1 - D)^2 12 m_W^2 m_Z^2 s_W^4}
 \Bigg\{ 18 (\frac{1}{D} - 1) (4 - 6 D + D^2) m_W^2 m_Z^2}\no
 &&\times A_0(m_W^2)
 B_0(p^2; 0, 0) + (\frac{1}{D} - 1) (4 - 6 D + D^2)
 \Big[ (m_W^2 - m_Z^2)^2 + 9 m_W^4 \Big] A_0(m_Z^2)
 B_0(p^2; 0, 0) \no
 &&+ \frac{1}{2} (1 - D) m_Z^2
 \Big[16 m_W^4 + 4 m_W^2 m_Z^2 - 2 m_Z^4 +
 10 (7 - D) m_W^2 p^2 - (7 - D) m_Z^2 p^2 \Big]
  B_0^2(p^2; 0, 0) \no
 && + 36 (1 - D) m_W^2 (m_W^2 - m_Z^2) (m_W^2 + p^2) B_0(p^2; 0, 0)
 B_0(p^2; 0, m_W^2) - 36 (1 - D) m_W^4 (m_W^2 \no
 &&+ m_Z^2 + p^2) B_0(p^2; 0,
 0) B_0(p^2; m_W^2, m_Z^2) + \frac{9}{2} \frac{m_Z^2}{p^2} \Big[ (4 - D)
 m_W^4 - (3 - D) m_W^2 m_Z^2 + (\frac{1}{D} - 1) \no
 && \times (16 - 37 D + 20 D^2 - 4
 D^3) m_W^2 p^2 - 2 (3 - 2 D) (3 - D) p^4 \Big] T_{13'4'}(m_W^2)
 - \frac{1}{2 m_Z^4 p^2} \bigg[ 40 (5 \no
 &&- D) m_W^{10} - 4 (264 - 141 D + 20
 D^2) m_W^8 m_Z^2 + 6 (279 - 181 D + 28 D^2) m_W^6 m_Z^4 - 6 (191 \no
 &&- 134
 D + 22 D^2) m_W^4 m_Z^6 + 11 (31 - 23 D + 4 D^2) m_W^2 m_Z^8 - 11 (2 -
 D) m_Z^{10} - 80 (2 - D) (5 \no
 &&- D) m_W^8 p^2 + 8 (308 - 291 D + 91 D^2 - 10
 D^3) m_W^6 m_Z^2 p^2 - 4 (\frac{20}{D} + 543 - 570 D + 200 D^2 \no
 &&- 22
 D^3) m_W^4 m_Z^4 p^2 + 2 (\frac{8}{D} + 409 - 448 D + 174 D^2 - 22 D^3
 ) m_W^2 m_Z^6 p^2 - (\frac{1}{D} + 1) (8 + 5 D \no
 &&- 2 D^2) m_Z^8 p^2 + 40
 (3 - 2 D) (5 - D) m_W^6 p^4 - 4 (64 - 11 D) (3 - 2 D) m_W^4 m_Z^2 p^4 +
 22 (3 - 2 D) \no
 &&(5 - D) m_W^2 m_Z^4 p^4 \bigg] T_{13'4'}(m_Z^2) +
 \frac{m_W^2}{p^2} (m_W^2 - m_Z^2) \Big[9 m_W^2 + (33 - 37 D +
 13 D^2) p^2 \Big] T_{2'3'4'} \no
 &&- 5 (1 - D) (10 - D) m_W^2 (m_W^2
 - m_Z^2) p^2 T_{1' 2'3'4'} + 18 \frac{m_W^2}{p^2} (m_W^2 -
 m_Z^2) \Big[ (2 - D) (m_W^4 - p^4) \no
 &&- 4 (1 - D) m_W^2 p^2 \Big]
 T_{12'3'4'}(m_W^2) + 9 \frac{m_W^2}{p^2} (m_W^2 - m_Z^2) (m_W^2
 + p^2) \Big[(3 - 2 D) (m_W^4 + p^4) - 2 (5 \no
 &&- 4 D) m_W^2 p^2
 \Big] T_{112'3'4'}(m_W^2, m_W^2) + \frac{m_Z^2}{2 p^2} \Big[
 2 m_W^4 - 22 m_W^2 m_Z^2 + 11 m_Z^4 + 18 (\frac{16}{D} - 31 +
 18 D \no
 &&- 4 D^2) m_W^2 p^2 \Big] T_{23'4'}(m_W^2) + \frac{2 m_W^2
 }{m_Z^4 p^2} (m_W^2 - m_Z^2) \Big[ 20 m_W^8 - (71 - 20
 D) m_W^6 m_Z^2 + (51 - 20 D) \no
 &&m_W^4 m_Z^4 - 20 (5 - 2 D) m_W^6
 p^2 + (135 - 62 D) m_W^4 m_Z^2 p^2 - 22 (2 - D) m_W^2 m_Z^4 p^2
 + 20 (7 \no
 &&- 4 D) m_W^4 p^4 - (17 + 16 D) m_W^2 m_Z^2 p^4 - 3 (17
 - 14 D) m_Z^4 p^4 - 20 (3 - 2 D) m_W^2 p^6 + 11 (3 \no
 &&- 2 D) m_Z^2
 p^6 \Big] T_{1'23'4'}(m_W^2) + \frac{20 m_W^2}{m_Z^2 p^2}
 (m_W^2 - m_Z^2)^2 (m_W^2 - p^2) \Big( m_W^2 - (3 - 2 D) p^2 \Big)
 \Big[T_{1'1'3'4'} \no
 &&+ (m_W^2 - p^2) T_{1'1'23'4'}(m_W^2)\Big] - \frac{1}{m_Z^4
 p^2} \Big[40 m_W^{12} - 2 (91 - 20 D) m_W^{10} m_Z^2 + 4 (61 -
 20 D) m_W^8 m_Z^4 \no
 &&- 2 (28 - 9 D) m_W^6 m_Z^6 - 4 (28 - 11 D)
 m_W^4 m_Z^8 + 11 (7 - 2 D) m_W^2 m_Z^{10} - 11 m_Z^{12} - 40 (5
 - 2 D) \no
 && \times m_W^{10} p^2 + 2 (235 - 102 D) m_W^8 m_Z^2 p^2 - 2 (179
 - 84 D) m_W^6 m_Z^4 p^2 +2 (11 + 12 D) m_W^4 m_Z^6 p^2 \no
 &&- 22 (3 - 2 D) m_W^2 m_Z^8 p^2 +11 (3 - 2 D) m_Z^{10} p^2 + 40 (7 - 4
 D) m_W^8 p^4 - 2 (157 - 64 D) m_W^6 m_Z^2 p^4 \no
 &&- 4 (17 - 29 D)
 m_W^4 m_Z^4 p^4 + 66 (1 - D) m_W^2 m_Z^6 p^4 - 40 (3 - 2 D)
 m_W^6 p^6 + 62 (3 - 2 D) \no
 && \times m_W^4 m_Z^2 p^6 - 22 (3 - 2 D) m_W^2
 m_Z^4 p^6 \Big] T_{123'4'}(m_Z^2, m_W^2) + \frac{1}{2 m_Z^2
 p^2} \Big( 11 (m_W^2 - m_Z^2)^2 + 9 m_W^4 \Big)\no
 && \times \Big[ 2 m_W^8 -
 4 (3 - D) m_W^6 m_Z^2 + (19 - 8 D) m_W^4 m_Z^4 -2 (5 - 2 D)
 m_W^2 m_Z^6 + m_Z^8 - 2 (5 - 2 D) \no
 &&m_W^6 p^2 + 8 (2 - D) m_W^4
 m_Z^2 p^2 + 4 (6 - 5 D) m_W^2 m_Z^4 p^2 - 2 (3 - 2 D) m_Z^6 p^2
 + 2 (7 - 4 D) m_W^4 p^4 \no
 &&+ 12 (1 - D) m_W^2 m_Z^2 p^4 + m_Z^4 p^4
 - 2 (3 - 2 D) m_W^2 p^6 \Big] T_{1123'4'}(m_Z^2, m_Z^2, m_W^2)
 - \frac{1}{2 p^2} \Big[ 18 m_W^6 \no
 &&- 9 m_W^4 m_Z^2 - 2 (
 \frac{80}{D} - 183 + 122 D - 28 D^2 ) m_W^4 p^2 + 2
 (\frac{1}{D} - 1) (8 - 4 D + D^2) ( 2 m_W^2 \no
 &&- m_Z^2 ) m_Z^2 p^2 \Big] T_{23'4'}(m_Z^2) - 9 \frac{m_W^4}{p^2}
 \Big[ 2 (2 - D) m_W^4 - (7 - 4 D) m_W^2 m_Z^2 + (3 - 2 D) m_Z^4
 - 8 (1 \no
 &&- D) m_W^2 p^2 - (3 - 2 D) m_Z^2 p^2 - 2 (2 - D) p^4
 \Big] T_{123'4'}(m_W^2, m_Z^2) - \frac{9 m_W^4}{2 p^2} \Big[ 2
 (3 - 2 D) m_W^6 \no
 &&- (13 - 8 D) m_W^4 m_Z^2 +4 (2 - D) m_W^2 m_Z^4
 - m_Z^6 - 2 (7 - 6 D) m_W^4 p^2 - 6 (3 - 2 D) m_W^2 m_Z^2 p^2 \no
 &&+ 4 (2 - D) m_Z^4 p^2 - 2 (7 - 6 D) m_W^2 p^4 - (13 - 8 D) m_Z^2
 p^4 + 2 (3 - 2 D) p^6 \Big] \no
 && \times T_{1123'4'}(m_W^2, m_W^2, m_Z^2) -
 18 (\frac{1}{D} - 1) (4 - 8 D + D^2) m_W^2 m_Z^2 (m_W^2 + p^2)
 T_{1'2'34'}(m_W^2) - (\frac{1}{D} \no
 && - 1) (4 - 8 D + D^2) (m_Z^2 +
 p^2) \Big( (m_W^2 - m_Z^2)^2 + 9 m_W^4 \Big) T_{1'2'34'}(m_Z^2)
 - 8 (1 - D) m_W^2 (m_W^2 \no
 &&- m_Z^2) p^4 T_{1'2'3'4'5'} - 72 (1 -
 D) m_W^4 (m_W^2 - m_Z^2) p^2 T_{12'3'4'5'}(m_W^2) + 72 (1 - D)
 m_W^4 (m_W^2 m_Z^2 \no
 &&+ m_W^2 p^2 + m_Z^2 p^2) T_{123'4'5'}(m_Z^2,
 m_W^2) + \frac{1}{2} (1 - D) (4 m_W^2 - m_Z^2) (2 m_W^2 +
 m_Z^2) \Big( 2 m_Z^4 + (8 \no
 &&- D) m_Z^2 p^2 + 2 p^4 \Big) T_{1'2'34'5'}(m_Z^2) \Bigg\} \;
 , \label{eq:w}
\end{eqnarray}
where the 't Hooft-Feynman gauge was used.
\chapter{Integral relations}

We list here some relations needed for the calculations
performed in this paper. The reduction formula for the integral
$Y_{2 3 4 5} ^{1 1}$ reads
\begin{eqnarray}
Y_{2 3 4 5} ^{1 1} &=& 2 (m_2^2 + p^2) Y_{2 3 4
5} ^{1} + A_0(m_3^2) A_0(m_5^2) - (m_2^2 - m_3^2) A_0(m_3^2)
B_0(p^2;m_4^2,m_5^2)\no
&&
+
(m_4^2 - p^2) \Big[T_{ 3 5 5'} + (m_4^2 - p^2) T_{ 3 4 5 5'}
\Big]
- (m_2^2 + p^2)^2 T_{ 2 3 4 5}
 + m_2^2 \Big[  T_{ 2 3 4} - T_{ 2 3 5} - (2 m_4^2 \no
 &&- m_5^2
- 2 p^2) T_{ 2 3 4 5} \Big]
+ m_2^2 (m_4^2 - p^2)
\Big[ T_{ 2 3 5 5'}
+ (m_4^2 - p^2) T_{ 2 3 4 5 5'} \Big]  \no
&&  + \frac{p^2}{1 - D} \bigg\{ T_{ 2 3 4} -
\Big[ A_0(m_2^2)+A_0(m_3^2) \Big] B_0(p^2;m_4^2,m_5^2)
 - (m_2^2 - m_3^2) \Big[ T_{ 2 4 5 5'}  -  T_{ 3 4 5 5'} \Big]
 \no
 &&
 -
( 2 m_2^2 + 2 m_3^2 - m_5^2)
 T_{ 2 3 4 5}
 + (m_2^2 - m_3^2)^2 T_{ 2 3 4 5 5'}
 \bigg\}  \no
&&  - \frac{D}{4(1 - D)} \bigg\{ Y_{2 3 4}^5 -
A_0(m_2^2)  A_0(m_3^2) - \Big[ A_0(m_2^2) + A_0(m_3^2) \Big]
\Big[ A_0(m_4^2)-A_0(m_5^2) \Big]  \no
&&
 - ( 2 m_2^2 + 2 m_3^2 + 2 m_4^2 - m_5^2 - 2 p^2)
 T_{ 2 3 4} + (2 m_2^2 + 2 m_3^2 + m_4^2 - m_5^2
-p^2) T_{ 2 3 5}  \no
&& - ( m_2^2 - m_3^2 - 2 m_4^2 + m_5^2 + 2 p^2 ) T_{ 2
4 5} + ( m_2^2 - m_3^2 + 2 m_4^2 - m_5^2  - 2 p^2 ) T_{
3 4 5} \no
&& + (m_2^2 - m_3^2 - m_4^2 + p^2) T_{ 2 5 5'}
- (m_2^2 - m_3^2 + m_4^2 - p^2) T_{ 3 5 5'} \no
&&  + \Big[ (m_4^2 - m_5^2 - p^2)^2 +
2 (m_2^2 + m_3^2) (2 m_4^2 - m_5^2 - 2 p^2)
  + (m_2^2 - m_3^2)^2 \Big] T_{ 2
3 4 5} \no
&& - \Big[ 2 (m_2^2 + m_3^2) (m_4^2 - p^2)
  + (m_2^2 - m_3^2)^2 \Big] T_{ 2
3 5 5'}
 - \Big[ (m_4^2 - p^2)^2
 -
2 (m_2^2 - m_3^2) (m_4^2 \no
&& -
p^2) \Big] T_{ 2 4 5 5'}
- \Big[ (m_4^2 - p^2)^2
  + 2 (m_2^2 - m_3^2) (m_4^2 - p^2)
\Big] T_{ 3 4 5 5'} - (m_2^2 - m_3^2)(m_4^2 \no
&& - p^2) \Big[ T_{ 2 5 5'
5'}
 -
T_{3 5 5' 5'} \Big] - \Big[ 2 (m_2^2 - m_3^2)^2 (m_4^2 - p^2)
 + 2 (m_2^2 + m_3^2) (m_4^2 - p^2)^2 \Big]
T_{ 2 3 4 5 5'} \no
&&+ (m_2^2 - m_3^2)^2 (m_4^2 - p^2)
 T_{ 2 3 5 5' 5'}
  - (m_2^2 - m_3^2) (m_4^2 - p^2)^2
\Big[ T_{ 2 4 5 5' 5'} - T_{ 3 4 5 5' 5'} \Big]  \no
&&  + (m_2^2 - m_3^2)^2 (m_4^2 - p^2)^2
 T_{ 2 3 4 5 5' 5'} \bigg\} \; .
\end{eqnarray}
{}From this relation the formula for the integral $Y_{2 3 4 5 5}
^{1 1}$ can be obtained by taking the derivative with respect to
$m_5 ^2$.

For the integral $Y_{2 3 4}^{1 1}$ we get
\begin{eqnarray}
 Y_{2 3 4} ^{1 1} &=& \frac{1}{3 (4 - 5 D) (4 - 3 D)} \Bigg\{
 \Big[(-4 + 5 D) D m_2^2 - (16 - 11 D) D m_3^2
 + (5 - 4 D) (4 - 3 D) \no
 && \times D m_4^2
 + (48 - 100 D + 65 D^2 - 12 D^3) p^2 \Big] A_0(m_2^2) A_0(m_3^2) +
 \Big[ (48 - 76 D + 29 D^2) m_2^2 \no
 &&- (4 - 5 D) D m_3^2 - (16 - 11 D) D m_4^2 -
 (4 - 5 D) D p^2 \Big] A_0(m_2^2) A_0(m_4^2) +
 \Big[ (-4 + 3 D) \no
 && \times D (1 + D) m_2^2 + (48 - 76 D + 29 D^2) m_3^2 -
 (4 - 5 D) D m_4^2 - D (16 - 17 D + 3 D^2) p^2 \Big] \no
 && \times A_0(m_3^2) A_0(m_4^2) +
 \Big[(16 - 36 D + 17 D^2) m_2^4 + 2 (4 - D) D m_2^2 m_3^2
 - 6 D^2 m_2^2 m_4^2 + 2 (8 \no
 &&- 5 D) D m_2^2 p^2 + 2 D (2 + D) m_3^2 p^2 -
 6 (2 - D) D m_4^2 p^2 - (16 - 20 D + 7 D^2) p^4 \Big] \no
 && \times A_0(m_3^2) B_0(p^2; 0, m_2^2)
 + \Big[ (-16 + 24 D - 11 D^2) m_2^4 -
 2 (4 - D) D m_2^2 m_3^2 + 6 D^2 m_2^2 m_4^2 \no
 &&+ 2 (4 - D) D m_2^2 p^2 -
 2 D (2 + D) m_3^2 p^2 + 6 (2 - D) D m_4^2 p^2 +
 (16 - 32 D + 13 D^2) p^4 \Big] \no
 && \times A_0(m_4^2) B_0(p^2; 0, m_2^2) -
 2 (m_3^2 - p^2) \Big[(8 - 20 D + 11 D^2) m_2^2 - (4 - 3 D) D m_3^2 -
 (2 \no
 &&- D) D m_4^2 + (2 - 3 D) (4 - 3 D) p^2 \Big]
 \Big[ A_0(m_2^2) - A_0(m_4^2) \Big] B_0(p^2; 0, m_3^2)
 - 2 \Big[ (16 - 20 D \no
 &&+ 3 D^2)
  m_2^2 m_4^2 - (12 - 11 D) D m_3^2 m_4^2 -
 (8 - 7 D) D m_4^4 + (8 - 4 D - 3 D^2) m_2^2 p^2 -
 (24 \no
 &&- 36 D
 + 11 D^2) m_3^2 p^2 + 4 (1 - D) (4 - D) m_4^2 p^2 -
 (16 - 28 D + 11 D^2) p^4\Big] A_0(m_2^2) \no
 && \times B_0(p^2; 0, m_4^2)
 + 2 \Big[ (16 - 20 D + 3 D^2) m_2^2 m_4^2
 - (12 - 11 D) D m_3^2 m_4^2 +
 (24 - 44 D \no
 &&+ 19 D^2) m_4^4 + (8 - 4 D - 3 D^2) m_2^2 p^2 -
 (24 - 36 D + 11 D^2) m_3^2 p^2 - 4 (8 - 5 D) (1 - D) \no
 && m_4^2 p^2 + (8 - 8 D
 + D^2) p^4 \Big] A_0(m_3^2) B_0(p^2; 0, m_4^2) -
 3 D^2 (m_3^2 - m_4^2) (m_2^2 - p^2)^2 \Big[ A_0(m_3^2) \no
 && - A_0(m_4^2) \Big]
 B'_0(p^2; 0, m_2^2)
 + 12 (1 - D) D (m_2^2 - m_3^2) (m_4^2 - p^2)^2 \Big[ A_0(m_2^2)
 - A_0(m_3^2) \Big] \no
 && \times B'_0(p^2; 0, m_4^2)
 + 2 \Big[ (8 - 9 D) (2 - D) m_2^4 - 4 (1 - D) (4 - D) m_2^2 m_3^2 +
 (6 - 5 D) D m_3^4 \no
 &&+ (24 - 56 D + 37 D^2 - 6 D^3) m_2^2 m_4^2 -
 (4 - 3 D) (1 - 2 D) D m_3^2 m_4^2 -
 (8 - 20 D + 19 D^2 \no
 &&- 6 D^3) m_2^2 p^2 -
 (4 - 3 D) (4 - 7 D + 2 D^2) m_3^2 p^2 \Big] T_{134'}(m_3^2,
 m_2^2) + 2 (m_2^2 - m_4^2) \no
 && \times \Big[ (8 - 20 D
 + 11 D^2) m_2^2 - (4 - 3 D) D m_3^2 -
 (2 - D) D m_4^2 + (2 - 3 D) (4 - 3 D) p^2 \Big] \no
 && \times T_{134'}(m_4^2, m_2^2)
 + \Big[ (-4 + 3 D) (1 - D) (4 + D) m_2^2 m_3^2
 - (8 - 5 D) D m_3^4 + (1 - D) (16 \no
 &&- 20 D + 3 D^2)
 m_2^2 m_4^2 + 2 (4 - D) D m_3^2 m_4^2 -
 3 D^2 m_4^4 - (4 - 3 D) (1 - D) (4 - D) m_3^2 p^2 \no
 &&+ (1 - D)
 (16 - 4 D - 3 D^2) m_4^2 p^2 \Big] T_{134'}(m_4^2,
 m_3^2) + \Big[(-4 + 5 D) D m_2^4 + 8 (3 - 2 D) (2 \no
 &&- D) m_2^2 m_3^2
 - (4 - 5 D) D m_3^4 - 4 (4 - 5 D) (3 - D) m_2^2 m_4^2 -
 2 (22 - 17 D) D m_3^2 m_4^2 \no
 &&- (4 - 5 D) D m_4^4
 - 2 (22 - 17 D) D m_2^2 p^2 - 4 (4 - 5 D) (3 - D) m_3^2 p^2 +
 8 (3 - 2 D) (2 \no
 &&- D) m_4^2 p^2 - (4
 - 5 D) D p^4 \Big] T_{234}(m_4^2, m_3^2, m_2^2) +
 2 \Big[ (16 - 20 D + 3 D^2) m_2^4 m_4^2 - 8 (1 \no
 &&- D) (2 + D)
 m_2^2 m_3^2 m_4^2 +
 (12 - 11 D) D m_3^4 m_4^2 + (12 - 13 D) (2 - D) m_2^2 m_4^4 +
 (14 \no
 &&- 13 D) D m_3^2 m_4^4 + (8 - 4 D - 3 D^2) m_2^4 p^2 -
 8 (1 - D) (4 - D) m_2^2 m_3^2 p^2 + (24 - 36 D \no
 &&+ 11 D^2) m_3^4 p^2
 - 8 (1 - D) (4 - D) m_2^2 m_4^2 p^2
 - 8 (1 - D) (2 + D) m_3^2 m_4^2 p^2 +
 (8 - 2 D \no
 &&- 5 D^2) m_2^2 p^4
 + (16 - 22 D + 5 D^2) m_3^2 p^4 \Big] T_{1'234}(m_4^2, m_3^2,
 m_2^2) - 12 (1 - D) D (m_2^2 \no
 &&- m_3^2)^2 (m_4^2 - p^2)^2
 T_{1'1'234}(m_4^2, m_3^2, m_2^2) +
 2 (m_2^2 - m_4^2) (m_3^2 - p^2)
 \Big[(8 - 20 D + 11 D^2) \no
 && \times m_2^2 - (4 - 3 D)
 D m_3^2 - (2 - D) D m_4^2 +
 (2 - 3 D) (4 - 3 D) p^2 \Big] T_{1'234}(m_3^2, m_4^2, m_2^2) \no
 &&- 2 \Big[ (8 - 12 D + 7 D^2) m_2^4 m_3^2 + (4 - D) D m_2^2 m_3^4 -
 (4 - 7 D) (2 - D) m_2^4 m_4^2 - 2 D (2 \no
 && + D) m_2^2 m_3^2 m_4^2
 + 3 D^2 m_2^2 m_4^4 - 2 D (2 + D) m_2^2 m_3^2 p^2
 + D (2 + D) m_3^4 p^2 - 2 (4 - D) \no
 && \times D m_2^2 m_4^2 p^2
 - 2 (4 - D) D m_3^2 m_4^2 p^2 +
 3 (2 - D) D m_4^4 p^2 - (8 - 16 D + 5 D^2) m_3^2 p^4 \no
 &&+ (8 - 10 D
 + 5 D^2) m_4^2 p^4 \Big] T_{1'234}(m_2^2, m_4^2, m_3^2) \no
 &&+ 3 D^2 (m_3^2 - m_4^2)^2 (m_2^2 - p^2)^2 T_{1'1'234}(m_2^2,
 m_4^2, m_3^2) \Bigg\} \; .
  \label{eq:ap11Y234}
\end{eqnarray}
The integral $Y_{2 3 4}^{1 5}$ is expressible through $Y_{2 3
4}^{1 1}$ and integrals of simpler structure via
\begin{eqnarray}
 Y_{2 3 4}^{1 5} &=& -\frac{1}{2} Y_{2 3 4}^{1 1} + \frac{1}{2}
 \bigg\{ Y_{2 3}^{1} + Y_{2 4}^{1} + Y_{3 4}^{1} + (m_2^2 +
 m_3^2 + m_4^2 + p^2) Y_{2 3 4}^{1} + (m_2^2 - p^2) A_0(m_2^2)
 \no
 && \times \Big[ A_0(m_3^2) - A_0(m_4^2) \Big]
 - (m_3^2 - m_4^2) A_0(m_3^2)
 A_0(m_4^2) - (m_2^2 - p^2) (m_3^2 - m_4^2) T_{234} \bigg\} \; . \no
 &&
\end{eqnarray}

The expression for the integral $T_{2345'5'}(m^2, M^2, M^2)$
in terms of integrals with fewer propagators reads
\begin{eqnarray}
\lefteqn{
T_{2345'5'}(m^2, M^2, M^2) =
\frac{1}{ D (m^2 - M^2)^2 (M^2 - p^2)^2 } \bigg\{
(2 - D) \Big[2 m^2 - (4 - D) M^2 + (2  } \hspace{0.5 cm} \no
&&- D) p^2 \Big] A_0(m^2) A_0(M^2) + 2 (2 - D)
(M^2 - p^2) A_0^2(M^2) - 2 (2 - D)
(m^2 -  p^2)^2 A_0(M^2) \no
&& \times B_0(p^2; 0, m^2) - \Big[D m^2 M^2 - (8 - 3 D) M^4 + (4
- D) m^2 p^2 + (4 - 3 D) M^2 p^2 \Big] A_0(m^2) \no
&& \times B_0(p^2; 0, M^2) +
\Big[D m^2 M^2 - (4 - D) \Big(M^4 + (m^2 - M^2) p^2 \Big) +
2 (2 - D) p^4 \Big] A_0(M^2) \no
&& \times B_0(p^2; 0, M^2) +
D (m^2 - M^2) (M^2 - p^2)^2 \Big[ A_0(m^2) - A_0(M^2) \Big]
B'_0(p^2; 0, M^2) - (2 \no
&&- D) \Big[ (2 + D) m^2 M^2 - (4 - D) M^4 + (2 - D) m^2 p^2
- D M^2 p^2 \Big] T_{235'}(m^2, M^2) \no
&&- 4 M^2 (m^2 - p^2) T_{235'}(M^2, M^2)
+ (8 - 3 D) (m^2 - M^2) (M^2 - p^2) T_{234}(m^2, M^2, M^2) \no
&&+ \Big[ D (m^4 M^2 + M^2 p^4) - (4 - D) \Big( m^2 (M^4
- p^4) - p^2 (m^4 - M^4) - 2 M^6 \Big) \no
&&- 8 m^2 M^2 p^2 \Big] T_{2345'}(m^2, M^2, M^2)
 - 4 M^2 (m^2 - p^2)^2 T_{1'234}(m^2, M^2, M^2)
    \bigg\} \; .
\end{eqnarray}

\end{document}